\documentclass[showpacs,floatfix,twocolumn]{revtex4-1}
\usepackage{graphicx,psfrag,amsmath,amssymb,amsfonts,latexsym,color,epsf,dcolumn,graphpap}

\usepackage[labelsep=period]{caption}
\usepackage{subcaption}
\usepackage{comment}
\usepackage{float}
\usepackage{graphicx}
\usepackage{cancel}
\usepackage[utf8]{inputenc}                 
\usepackage[T1]{fontenc}                    

\usepackage{graphicx}                       
\usepackage{amsthm, amsmath, amssymb}       
\usepackage[loose,nice]{units}              
\usepackage{lipsum}                         

\usepackage{bookmark}                       
\usepackage{fancyhdr}                       
\usepackage{emptypage}                      
\usepackage{color}
\usepackage{ulem}
\usepackage{physics}                        

\usepackage[svgnames]{xcolor}
\usepackage{hyperref}                       
\usepackage{cancel}
\usepackage{bm}
\hypersetup{
    colorlinks=true,
    linkcolor=blue,
    filecolor=magenta,      
    urlcolor=cyan,
    citecolor=Green,
    hyperindex=true,
    pdfauthor={Jean Paul Louys Sanso},
    pdftitle={paperjp},
    }

\makeatletter
\def\thickhrulefill{\leavevmode \leaders \hrule height 1ex \hfill \kern \z@}

\usepackage{tikz-cd} 
\usepackage{tikz}
\usepackage{appendix}
\usetikzlibrary{shapes.geometric, arrows}

\definecolor{lime}{HTML}{A6CE39}
\DeclareRobustCommand{\orcidicon}{
	\begin{tikzpicture}
	\draw[lime, fill=lime] (0,0) 
	circle [radius=0.16] 
	node[white] {{\fontfamily{qag}\selectfont \tiny ID}};
	\draw[white, fill=white] (-0.0625,0.095) 
	circle [radius=0.007];
	\end{tikzpicture}
	\hspace{-2mm}
}

\foreach \x in {A, ..., Z}{%
	\expandafter\xdef\csname orcid\x\endcsname{\noexpand\href{https://orcid.org/\csname orcidauthor\x\endcsname}{\noexpand\orcidicon}}
}

\definecolor{red}{rgb}{1,0,0}
\definecolor{blue}{rgb}{0,0,1}
\definecolor{skyblue}{rgb}{0,0,.5}
\definecolor{green}{rgb}{0,1,0}
\definecolor{orange}{cmyk}{0,.4,1,0}

\begin{document}
\title{Photon Generation in Double Superconducting Cavities:  \\ Quantum Circuits Implementation}

\author{Jean Paul Louys Sansó$^{1,2}$\orcidA}
\author{Nicol\'as F.~Del Grosso$^{1,3}$\orcidN}
\author{Fernando C. Lombardo$^{1,3}$\orcidF}
\author{Paula I.~Villar$^{1,3}$ }

\affiliation{$^1$ Departamento de F\'\i sica {\it Juan Jos\'e
 Giambiagi}, FCEyN UBA, Facultad de Ciencias Exactas y Naturales,
 Ciudad Universitaria, Pabell\' on I, 1428 Buenos Aires, Argentina }
 \affiliation{$^2$ Institut für Experimentalphysik, Universität Innsbruck, Technikerstraße 25, 6020 Innsbruck, Austria}
 \affiliation{$^3$ IFIBA CONICET-UBA, Facultad de Ciencias Exactas y Naturales,
 Ciudad Universitaria, Pabell\' on I, 1428 Buenos Aires, Argentina }
 
\begin{abstract}
In this work, we studied photon generation due to the Dynamical Casimir Effect (DCE) in a one dimensional (1+1) double superconducting cavity. The cavity consists of two perfectly conducting mirrors and a dielectric membrane of infinitesimal depth that effectively couples  two cavities. The total length of the double cavity $L$, the difference in length between the two cavities $\Delta L$, and the electric susceptibility $\chi$ and conductivity $v$ of the dielectric membrane are tunable parameters. All four parameters are treated as independent and are allowed to be tuned at the same time, even  with different frequencies. We analyzed the cavity's energy spectra under different conditions, finding a transition between two distinct regimes that is accurately described by $k_c=\sqrt{v/\chi}$. In particular, a lowest energy mode is forbidden in one of the regimes while it is allowed in the other. We compared analytical approximations obtained through the Multiple Scale Analysis method with exact numeric solutions, obtaining the typical results when $\chi$ is not being tuned. However, when the susceptibility $\chi$ is tuned, different behaviours (such as oscillations in the number of photons of a cavity prepared in a vacuum state) might arise if the frequencies and amplitudes of all parameters are adequate. These oscillations can be considered as adiabatic shortcuts where all generated photons are eventually destroyed. Finally, we present an equivalent quantum circuit that would allow to experimentally simulate the DCE under the studied conditions.

\end{abstract}

\date{today}
\maketitle
\noindent 

\maketitle

\section{Introduction}\label{sec:intro}

The Dynamical Casimir Effect (DCE) is a purely quantum effect that has so far escaped experimental detection in a cavity, due to the need of moving mechanical mirrors at relativistic speeds to produce observable results \cite{moore1970,Dodonov96,plunien2000dynamical,crocce2001resonant,dewitt1975quantum,fulling1976radiation}. Many alternatives to the cavity with moving mirrors have been explored in the last decades, finding systems with analogous time-dependent boundary conditions \cite{johansson2009dynamical, wustmann2013parametric, Dodonov96, schaller2002dynamical, kim2006detectability, lambrecht1996motion, sassaroli1994photon}. This allowed the experimental observation of the DCE in one of the most promising systems to study it: superconducting quantum circuits \cite{wilson2011observation, johansson2013nonclassical}. Through the use of Superconducting Quantum Interference Devices (SQUIDs), effective lengths and other properties of the cavity can be tuned by applying external magnetic fields \cite{wustmann2013parametric, wilson2011observation, crocce2004model, nation2012colloquium, lozovik1995parametric, uhlmann2004resonant, johansson2013nonclassical, johansson2010dynamical, C.Braggio_2005, SEGEV2007202}. 

In this work we consider a massless scalar field in a one-dimensional double cavity in vacuum with a dielectric membrane that couples the left and right cavities. Including a dielectric membrane allows, under the proper driving conditions, an exponential growth in the number of photons. In a previous work \cite{velasco2022photon}, some of us have studied a similar setup where only the position 
of the membrane was tunable, but the physical characteristics of the membrane were fixed. 
In this sense, in the present work we give a much more complete description of the system 
and we obtain many more results than the previous case, which, obviously, is obtained from this 
more general case in the corresponding limits (see below). Optomechanical cavities with dielectric membranes have also shown great controllability of the parameters of the system  \cite{PhysRevA.99.023851}.  We consider all four cavity parameters as independently tunable in time: the length of each cavity and the electric conductivity and susceptibility of the dielectric membrane. We study the effects of the individual tuning of the parameters, and compare them to those resulting from tuning more than one parameter at the same time. In studying multi-parameter tuning, we consider different relative amplitudes and phases.

In Section \ref{sec:model} we present the general model for our double cavity, along with the dynamic equations and time-dependent boundary conditions. We also show how to compute the photons produced by the DCE after the movement stops. Along Section \ref{sec:spectrum} we discuss in detail the energy spectrum of the double cavity. Knowing how the spectrum behaves is critical to understand photon generation due to the DCE. We consider the effects of the variation of every parameter of the cavity. We also distinguish between behaviours that arise in different regimes. Section \ref{sec:eigenfunc} contains the study of the eigenfunctions localization for an arbitrary double cavity. In Section \ref{sec: photon} we introduce the Multiple Scale Analysis (MSA) method as one way to obtain an analytic  approximation to the coupled set of linear equations of the problem. The MSA method is more extensively discussed in Appendix \ref{ap:MSA}. We compare those results with numerical solutions under the usual coupling conditions: parametric resonance, additive coupling and difference coupling. We additionally study the effect of a detuning from the desired driving frequency in parametric resonance.

In Section \ref{sec:circuits} we show an experimental setup in which all the studied effects could be replicated. For that, we propose to use superconducting elements in a quantum circuits architecture. This allows to replace the tuning of the actual parameters with the tuning of external magnetic fields and capacitances. Finally, Section \ref{sec:conclusiones} summarizes the results of this work.

\section{Cavity model}\label{sec:model}
\begin{figure}[H]
    \centering
    \includegraphics[width = .45 \textwidth]{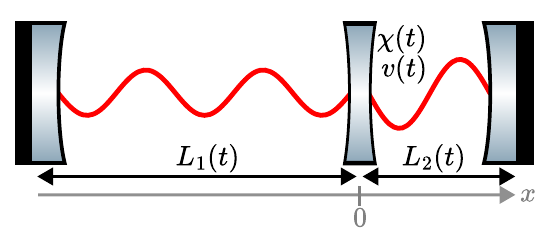}
    \caption{Schematic of the one-dimensional double cavity. All four time-dependent parameters can be tuned separately.}
    \label{fig: esquema_cavidad}
\end{figure}

We shall consider a massless scalar field $\phi$ confined in a one-dimensional double cavity in vacuum schematized in FIG. \ref{fig: esquema_cavidad}. The cavity consists of two perfectly conducting mirrors located at $x=-L_1(t)<0$ and $x=L_2(t)>0$, and a dielectric membrane fixed at $x=0$. The electric properties of the dielectric wall are determined by its electric susceptibility $\chi(t)$ and its electric conductivity $v(t)$. We consider the membrane as infinitely thin. The Lagrangian density of such cavity can be written as \cite{crocce2004model,Fosco2013}:
\begin{equation}
    \mathcal{L}=\frac{1}{2}[\epsilon(x,t)(\partial_t\phi)^2-(\partial_x\phi)^2-V(x,t)\phi^2], \label{eq:densidad_lagrangiana}
\end{equation}
where the electric permitivity $\epsilon$ and the electric potential $V$ are those of vacuum, except at the membrane ($x=0$):
\begin{align}
\epsilon(x,t)&=1+\chi(t)\delta(x), \label{eq:permitividad}\\ 
V(x,t)&=v(t)\delta(x). \label{eq:conductividad}
\end{align}

Everywhere but at $x=0$ the lagrangian density $\mathcal{L}$ in Eq. (\ref{eq:densidad_lagrangiana}) is the Klein-Gordon one \cite{peskin2018introduction}. Then $\phi$ satisfies the wave equation of a massless field:
\begin{equation}
    \square \phi =0, \label{eq: ec_ondas}
\end{equation}
and we impose the existence of the dielectric wall as a boundary condition for $\phi$. This condition can be obtained by integrating the Euler-Lagrange equation for the field and implies a jump discontinuity in the first spatial derivative of $\phi$:
\begin{align}
    \partial_x&\phi(t,0^+)-\partial_x\phi(t,0^-)=\lim_{d\to0}\int^d_{-d}\partial_x^2\phi~dx \notag \\ 
    &=\dot{\chi}(t)\partial_t\phi(t,0)+\chi(t)\partial_t^2\phi(t,0)+v(t)\phi(t,0).\label{eq:boundary4}
\end{align}
Note that the first term in the RHS of Eq. \eqref{eq:boundary4} vanishes for any fixed value of the electric susceptibility, unlike the other two terms. 

The remaining time-dependent boundary conditions are imposed by the perfectly conducting mirrors at both ends of the cavity, and by the continuity of the field at $x=0$:
\begin{align}
    \phi\left(t,x=-L_1(t)\right)&=0, \label{eq:boundary1}\\
    \phi\left(t,x=L_2(t)\right)&=0, \label{eq:boundary2}\\
    \phi(t,x=0^+)&=\phi(t,x=0^-).\label{eq:boundary3}
\end{align}

We will assume the cavity parameters are only tuned during a finite time $t_f$. The cavity will remain static at $t<0$ and $t\geq t_f$ in order to have well defined initial and final Fock states, required to compute the number of particles created by the DCE. In the following sections, we will refer to the static regions as the \textit{in} region ($t<0$) or the \textit{out} region ($t\geq t_f$).

\subsection{Eigenfunctions}
Because the boundary conditions are time-dependent, we consider an instantaneous spatial solution basis given by:
\begin{widetext}
\begin{equation}
\varPhi_n(t, x)\equiv\frac{1}{N_n}\begin{cases}
\sin(k_n(x+L_1)), &-L_1\leq x\leq0\\
-\frac{\sin(k_nL_1)}{\sin(k_nL_2)}\sin(k_n(x-L_2)), &\phantom{-L}0\leq x\leq L_2
\end{cases}, \label{eq:modos_espaciales_dinamicos}
\end{equation}
\end{widetext}
where $k_n$ are the instantaneous eigenfrequencies and $N_n$ are normalization constants. Here, the time dependency of the spatial functions $\varPhi_n$ is hidden through the time-dependent cavity parameters (both explicitly for the cavity lengths and implicitly for every parameter through $k_n$).
The multiplying fraction in the right cavity ($0\leq x\leq L_2$) spatial functions should be taken as a limit in the cases where the denominator is zero. It can be shown that if the denominator is equal to zero then the limit equals:
\begin{equation}
    \lim_{k\to k_n}-\frac{\sin(k L_1)}{\sin(k L_2)}=(-1)^{n_1+n_2},
\end{equation}
where $n_1$, $n_2\in \mathbb{Z}^+$ are such that:
\begin{equation}
    k_n=\frac{n_1\pi}{L_1}=\frac{n_2\pi}{L_2}. \label{eq:ks_caso1B}
\end{equation}

The spatial basis defined in Eq. \eqref{eq:modos_espaciales_dinamicos} is orthonormal with respect to the inner product \cite{Fosco2013}:
\begin{equation}
    (\varPhi_n, \varPhi_m)=\int^{L_2}_{-L_1}dx ~\varPhi_n(x)\varPhi_m^{*}(x)\left(1+\delta(x)\chi\right). \label{eq:producto_interno}
\end{equation}

The eigenfunctions $u_m$ consist of an expansion with time dependent coefficients $Q_n^{(m)}$ \cite{Dodonov96,law1995interaction}:
\begin{equation}
    u_m(t,x)=\sum_n Q_n^{(m)}(t)\varPhi_n(t,x).\label{eq:desarrollo_fourier}
\end{equation}
The set of instantaneous eigenfunctions is orthonormal with respect to the Klein-Gordon modified inner product \cite{Miladinovic2011}:
\begin{widetext}
\begin{equation}
    (f(t,x),g(t,x))_{KG}= i\int_{-L_1}^{L_2} dx~\left[\dot{f}(t,x)g^*(t,x)-f(t,x)\dot{g}^*(t,x)\right]. \label{eq:kg_inner} 
\end{equation}
\end{widetext}

At the static \textit{in} and \textit{out} regions, the spatial functions defined in Eq. \eqref{eq:modos_espaciales_dinamicos} are time-independent, and the eigenfunctions are of the form \cite{Miladinovic2011}:
\begin{equation}
u_n(t,x)=\frac{e^{-ik_n t}}{\sqrt{2k_n}}\varPhi_n(x). \label{eq:modos_estaticos}
\end{equation}
We must note that the eigenfunctions of the \textit{in} region are different than those of the \textit{out} regions. They are, however, connected by a Bogoliubov transformation as we shall discuss in section \ref{subsec:Q_coefficients}.

The continuity of the $u_m$ eigenfunctions at $t=0$ imposes the initial conditions for the $Q_n^{(m)}$ coefficients in Eq. \eqref{eq:desarrollo_fourier}:
\begin{align}
    Q_n^{(m)}(0)&=\frac{1}{\sqrt{2k_n}}\delta_{nm}, \label{eq:cond_inicial_Q}\\
    \dot{Q}_n^{(m)}(0)&=-i\sqrt{\frac{k_m}{2}}\delta_{nm}\label{eq:cond_inicial_Qdot}. 
\end{align}

\subsection{Time-dependent coefficients} \label{subsec:Q_coefficients}

By substituting Eq. \eqref{eq:desarrollo_fourier} in Eq. \eqref{eq: ec_ondas}, we obtain a set of coupled linear differential equations for the time-dependent coefficients $Q_n^{(m)}$:
\begin{align}
\ddot{Q}_l^{(m)}+ k_l^2 Q_l^{(m)}=-\sum_n Q_n^{(m)}\lambda_{nl}+2\dot{Q}_n^{(m)}\theta_{nl},\label{eq:ecuacion_Q}
\end{align}
where $\theta_{nl}$ and $\lambda_{nl}$ are inner products as defined in Eq. \eqref{eq:producto_interno}:
\begin{align}
    \theta_{nl}&\equiv(\partial_t\varPhi_n, \varPhi_l), \label{eq:theta} \\
    \lambda_{nl}&\equiv(\partial_t^2\varPhi_n, \varPhi_l). \label{eq:lambda}
\end{align}

In both static regions, we can expand the scalar field in terms of creation ($\hat{a}^{\dagger}_n$) and annihilation ($\hat{a}_n$) operators of particles with defined momentum: 
\begin{equation}
    \hat{\phi}(t,x)=\sum_n \hat{a}_n u_n(t,x)+\hat{a}^{\dagger}_n u_n^*(t,x) \label{eq:desarrollo_campo_in}. 
\end{equation}
The \textit{in}/\textit{out} operators are connected by a Bogoliubov transformation \cite{Dodonov96,Birrel82}:
\begin{equation}
    \hat{a}_n^{\textit{out}}=\sum_m \left(\alpha_{nm}\hat{a}_m^{\textit{in}}+\beta_{nm}\hat{a}^{\dagger\textit{in}}_m\right),\label{eq :bogoliubov}
\end{equation}
where $\alpha_{nm}$ and $\beta_{nm}$  coefficients can be obtained by evolving the initial $u_m^{\textit{in}}$ modes and projecting them against the final modes \cite{velasco2022photon}.

The RHS of Eq. \eqref{eq:ecuacion_Q} vanishes if the cavity is static, decoupling all modes and yielding harmonic oscillations for $t\geq t_f$:
\begin{equation}
    Q_n^{(m)}(t\geq t_f)=A_n^{(m)}e^{ik_n t}+B_n^{(m)}e^{-ik_n t}. \label{eq:oscilador_armonico}
\end{equation}
The $A_n^{(m)}$ and $B_n^{(m)}$ coefficients are related to the Bogoliubov coefficients of Eq. \eqref{eq :bogoliubov} through the eigenfrequencies \cite{velasco2022photon}:
\begin{align}
    \alpha_{nm}&=\sqrt{2k_n}B_n^{(m)},\label{eq:bogo1}\\
    \beta_{nm}&=\sqrt{2k_n}A_n^{(m)}\label{eq:bogo2}.
\end{align}

By introducing Eq. \eqref{eq:bogo1} and \eqref{eq:bogo2} in Eq. \eqref{eq :bogoliubov} we can compute the expected number of photons $\expval{N_n}$ with eigenfrequency $k_n$, for an initial state $\ket{\textit{in}}$:
\begin{align}
\expval{N_n}&=\bra{\textit{in}}\hat{N}_n^{\textit{out}}\ket{\textit{in}}=\bra{\textit{in}}\hat{a}^{\dagger\textit{out}}_n\hat{a}^{\textit{out}}_n\ket{\textit{in}}\notag\\
&= 2k_n \sum_m \left(\left(1+N_m^{(0)}\right)\left|A_n^{(m)}\right|^2+N_m^{(0)}\left|B_n^{(m)}\right|^2\right).\label{eq:numero_fotones}
\end{align}

In Eq. \eqref{eq:numero_fotones}, $N_m^{(0)}$ is the number of photons with $k_m$ frequency in the initial state. Eq. \eqref{eq:numero_fotones} shows that photon generation due to the DCE is possible even when the cavity is prepared in a vacuum state, given that the cavity parameters are tuned such that at least one of the $A_n^{(m)}$ coefficients are not zero. Photon generation rate and functional form will depend on the time-dependent $A_n^{(m)}$ and $B_n^{(m)}$ coefficients.

\section{Energy spectrum} \label{sec:spectrum}
Understanding the cavity's energy spectrum and how its parameters change it is vital to solving the photon generation problem. In this section, we will discuss the energy spectrum extensively.

By substituting the general solutions of Eq. \eqref{eq:modos_espaciales_dinamicos} in Eq. \eqref{eq:boundary4} we obtain a transcendental equation linking the (instantaneous) eigenfrequencies $k$ with all four tunable parameters of the cavity:
\begin{equation}
    \frac{2\sin(kL)}{k^2\chi+ik\dot{\chi}-v}k-\cos(k\Delta L)+\cos(kL) =0\label{eq:trascendente_estatica}.
\end{equation}
In Eq. \eqref{eq:trascendente_estatica}, $L=L_1+L_2$ is the total length of the cavity and $\Delta L=L_1-L_2$ its asymmetry.

This transcendental equation is also time-dependent if computed out of the static regions, and allows the instantaneous eigenfrequencies to acquire a non zero imaginary part. This should be interpreted as a mode whose amplitude instantly increases or decreases exponentially. If the region is static or if the susceptibility remains at a fixed value, eigenfrequencies and instantaneous eigenfrequencies are strictly real. In this last case, the LHS of Eq. \eqref{eq:trascendente_estatica} has a singularity in $k=k_c>0$ if $\chi, ~v\neq0$, where we defined the critical eigenfrequency or critical energy $k_c$ as:
\begin{equation}
    k_c=\sqrt{\frac{v}{\chi}}. \label{eq:k_critico}
\end{equation}

We will discuss the energy spectra of general configurations with no complex values, and briefly show the shift introduced by having a time-dependent electric susceptibility.

\subsection{Real solutions ($\dot{\chi}=0$)}
In order to study the spectrum of the cavity, we fix the value of $L$ and define four dimensionless parameters:
\begin{align}
    a_1&\equiv kL,\\
    a_2&\equiv\frac{\Delta L}{L},\\
    a_3&\equiv\frac{\chi}{L},\\
    a_4&\equiv vL.
\end{align}
These combinations of parameters allow us to rewrite Eq. \eqref{eq:trascendente_estatica} as:  
\begin{equation}
    \frac{2\sin(a_1)}{a_1a_3-a_4/a_1}-\cos(a_1a_2)+\cos(a_1)=0. \label{eq:trascendente adimensional}
\end{equation}

First, we shall consider a transparent dielectric wall (or, equivalently, its absence) by choosing $\chi=v=0$. In this scenario, Eq. \eqref{eq:trascendente adimensional} admits analytical solutions, yielding evenly spaced energy levels as one would expect from the boundary conditions of two perfectly conducting mirrors. The spectrum is completely degenerate in the dimensionless asymmetry $a_2$ of the cavity. This result is also expected, as $\Delta L$ loses its physical meaning once the dielectric membrane is removed.

\begin{figure}[H]
    \centering
    \includegraphics[width=0.49\textwidth]{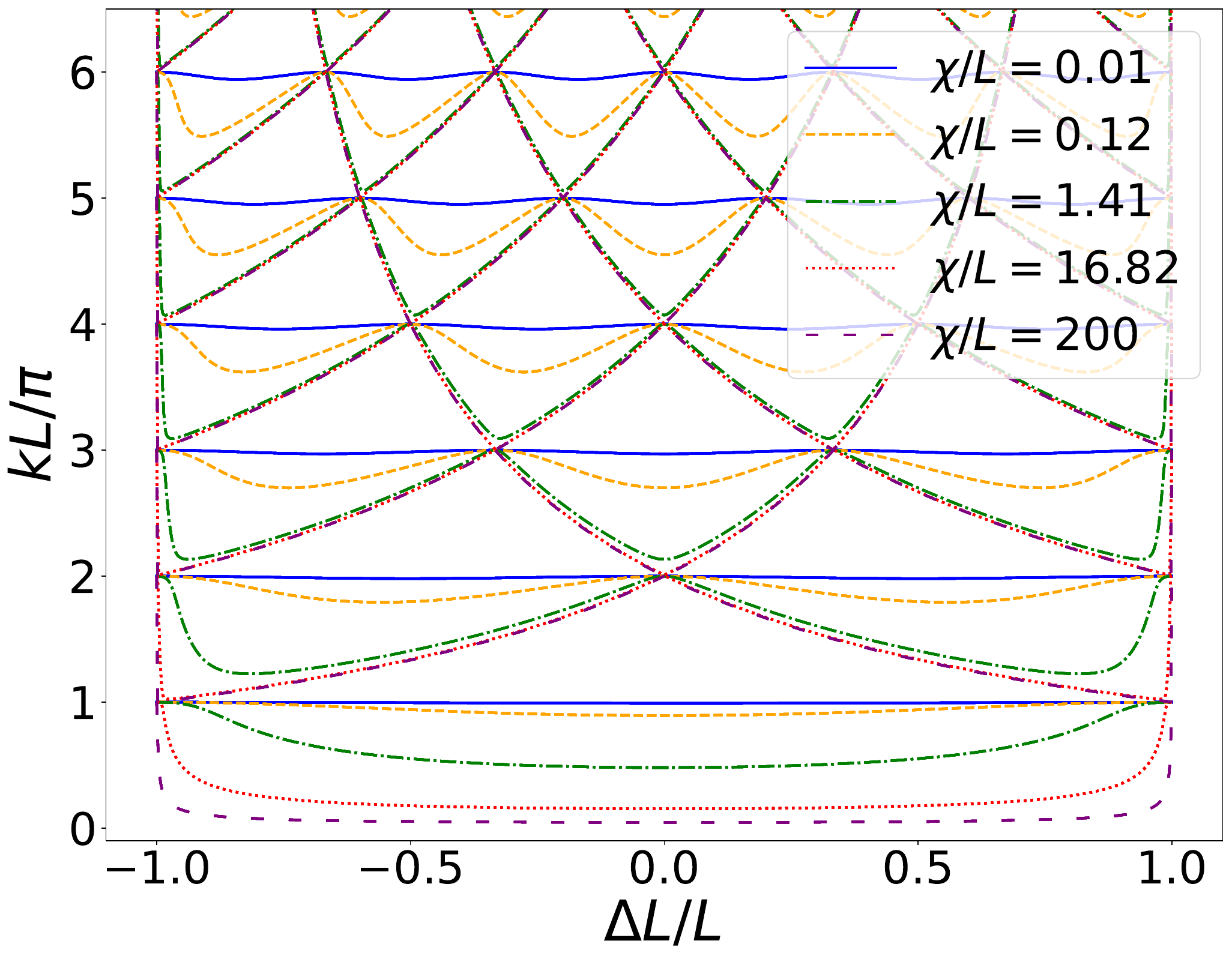}
    \caption{Energy spectrum of the cavity for $v=0$ and different values of $\chi$, as a function of the adimensional asymmetry $\Delta L/L$. }
    \label{fig:ch2_standardDCE_variandochi}
\end{figure}

\begin{figure}[H]
    \centering
    \includegraphics[width=0.49\textwidth]{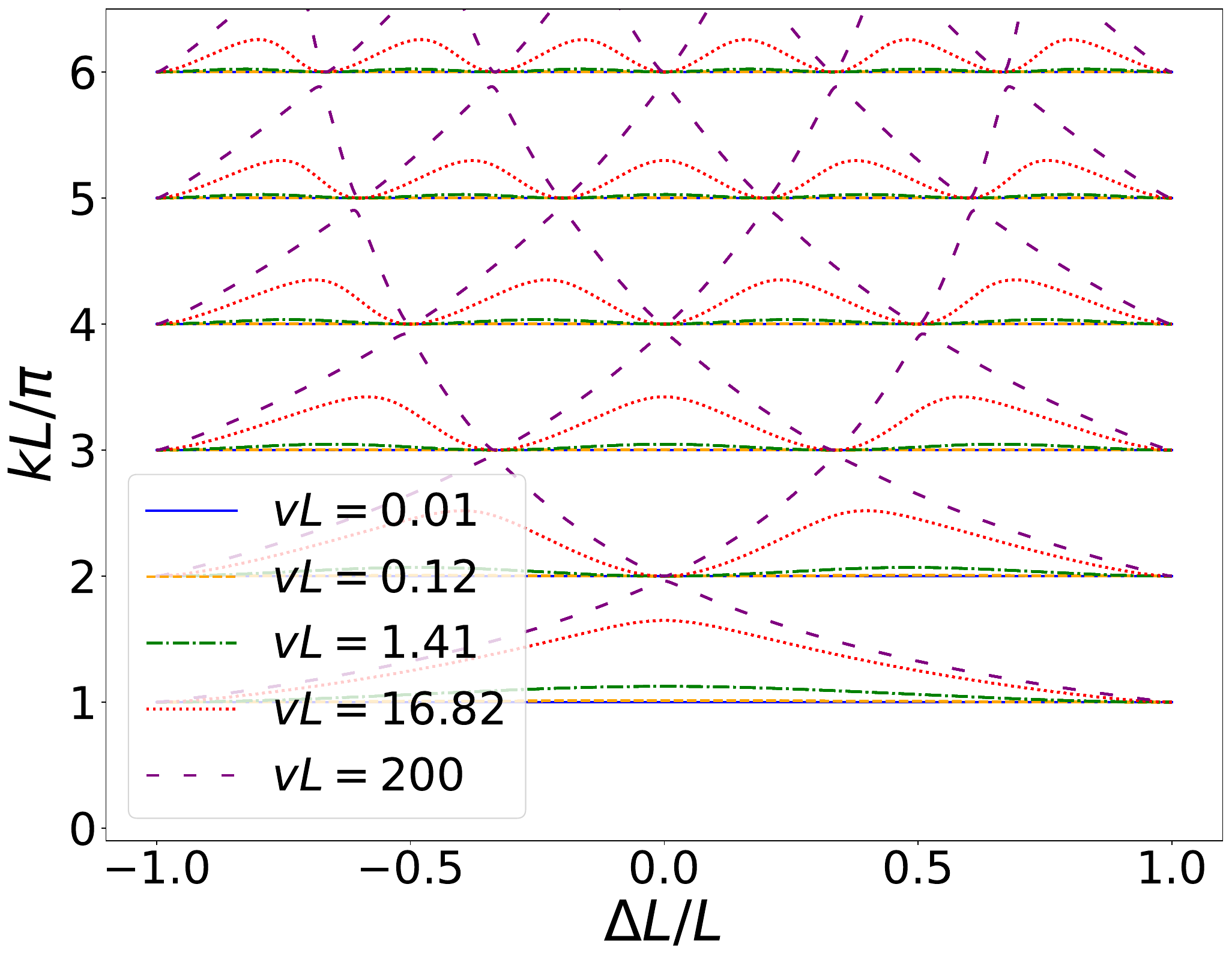}
    \caption{Energy spectrum of the cavity for $\chi=0$ and different values of $v$, as a function of the adimensional asymmetry $\Delta L/L$. }
    \label{fig:ch2_standardDCE_variandov}
\end{figure}

If $\chi\neq0$ or $v\neq0$, then the energy degeneracy in the asymmetry is partially broken. We show the energy spectra as a function of the asymmetry of two different regimes, one with $v=0$ (FIG. \ref{fig:ch2_standardDCE_variandochi}) and other with $\chi=0$ (FIG. \ref{fig:ch2_standardDCE_variandov}).

As expected, the dielectric membrane modifies the energy levels of the cavity. In FIG. \ref{fig:ch2_standardDCE_variandochi}, we can see that energy levels are lowered  up to $\pi/L$ by increasing the electric susceptibility $\chi$. For sufficiently high values of $\chi$, a low energy ($k\ll \pi/L$) level appears. This energy level only disappears in this regime when taking the  $\chi\to\infty$ limit. On the other hand, FIG. \ref{fig:ch2_standardDCE_variandov} shows a similar but opposite behaviour: energy levels are increased up to $\pi/L$ by increasing the electric conductivity $v$, and the existence of low energy ($k<\pi/L$) levels is completely prohibited. A similar case has been studied in \cite{crocce2004model}. In both FIG. \ref{fig:ch2_standardDCE_variandochi} and FIG. \ref{fig:ch2_standardDCE_variandov}, eigenfrequencies experience an avoided crossing. The gaps can be arbitrarily reduced by increasing $\chi$ in FIG. \ref{fig:ch2_standardDCE_variandochi} or $v$ in FIG. \ref{fig:ch2_standardDCE_variandov}, but only vanish for $v\to\infty$ or $\chi\to\infty$.

\begin{figure}[H]
    \centering
    \includegraphics[width=0.49\textwidth]{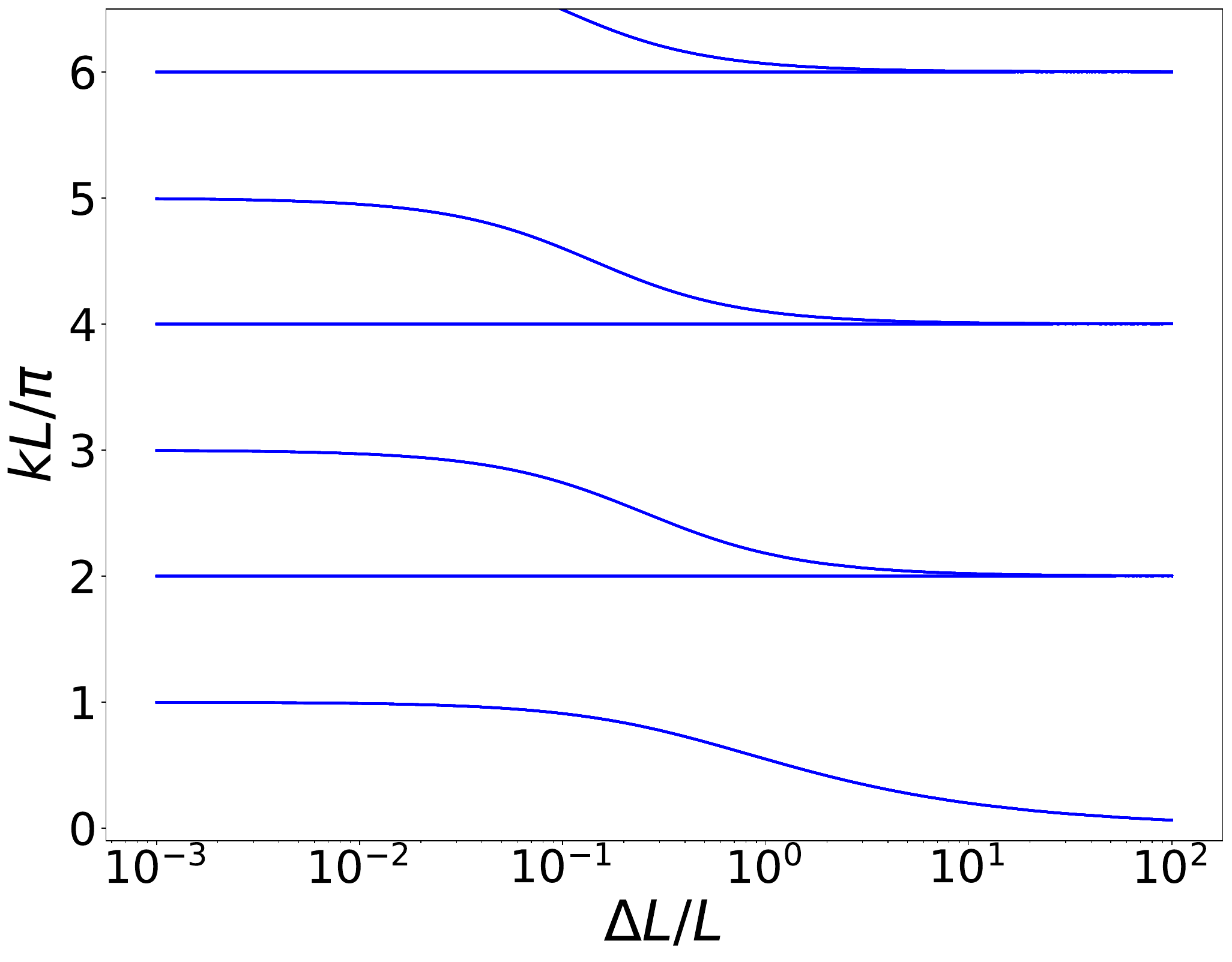}
    \caption{Energy spectrum of a perfectly symmetrical cavity ($\Delta L=0$) with $v=0$, as a function of the adimensional parameter $\chi/L$. Energy bands merge as the system reaches the $\chi\to\infty$ limit, resulting in an analogous solution to a mirror replacing the membrane.}
    \label{fig:merged}
\end{figure}

It can be shown that the same results arise from taking the $v\to\infty$ or $\chi\to\infty$ limits. This is consistent with the fact that both limits impose boundary conditions equivalent to those of a third perfectly conducting mirror, located at $x=0$ and replacing the dielectric membrane \cite{Miladinovic2011}. However, the $\chi\to\infty$ limit would be experimentally reached way sooner than the $v\to\infty$ limit. This can be seen in the similarities (differences) of the green, purple, and red curves of FIG. \ref{fig:ch2_standardDCE_variandochi} (FIG. \ref{fig:ch2_standardDCE_variandov}). By taking any of the above limits in a perfectly symmetrical cavity energy modes density halves, as shown in FIG. \ref{fig:merged}. Levels with energy equal to an odd multiple of $\pi/L$ merge towards the even multiples of $\pi/L$ by decreasing (increasing) the energy level as $\chi$ ($v$) increases. The lowest energy mode vanishes for sufficiently high values of $\chi$, yielding the same solutions in both limits. 

\begin{widetext}

\begin{figure}[H]
    \centering
    \begin{subfigure}[h]{0.49\columnwidth}
         \includegraphics[width=\linewidth]{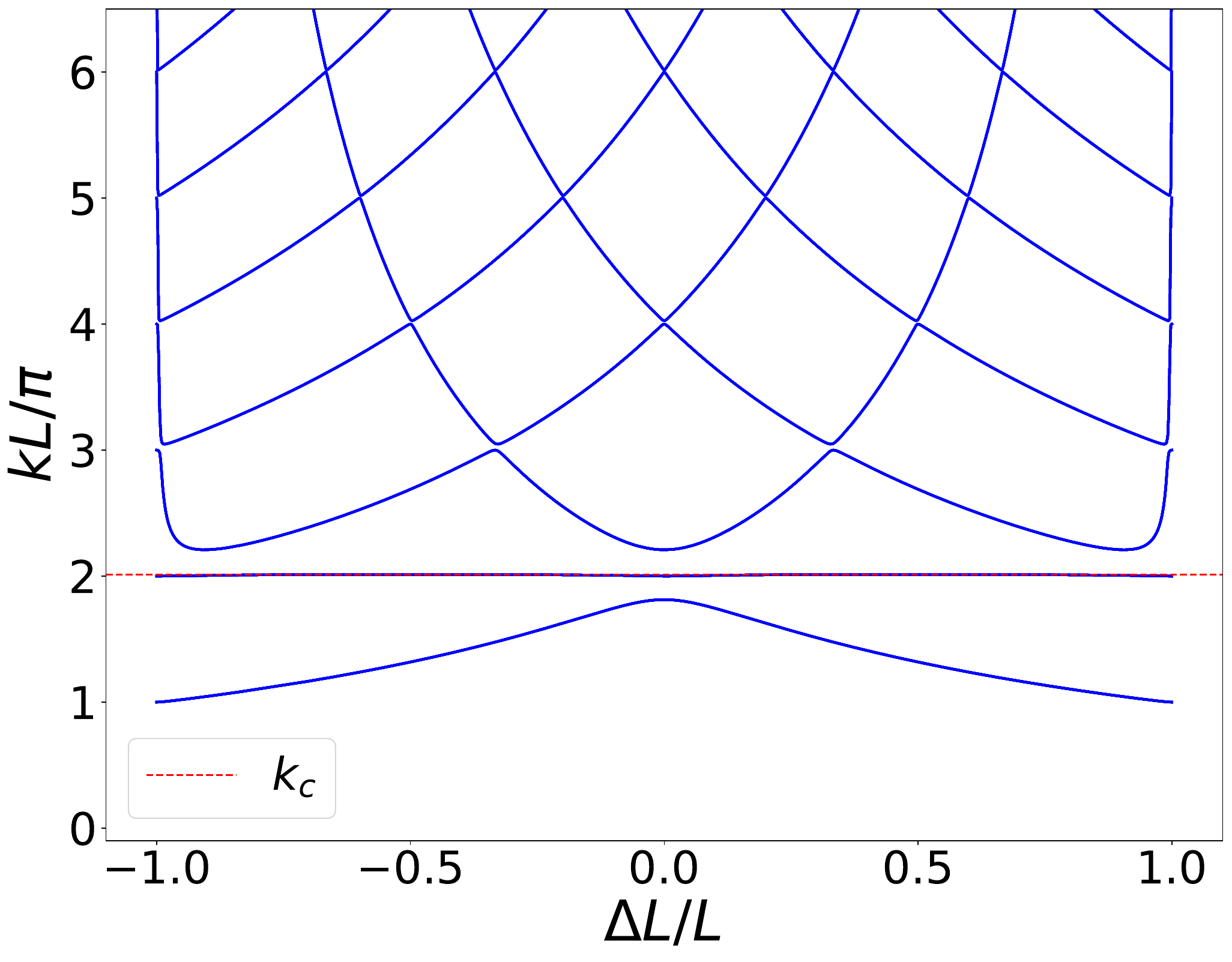}
         \caption{$\chi/L=5$, $vL=200$}
         \label{fig:ch2_generico_dL_critico}
     \end{subfigure}
     \\
     \begin{subfigure}[h]{0.49\columnwidth}
         \includegraphics[width=\linewidth]{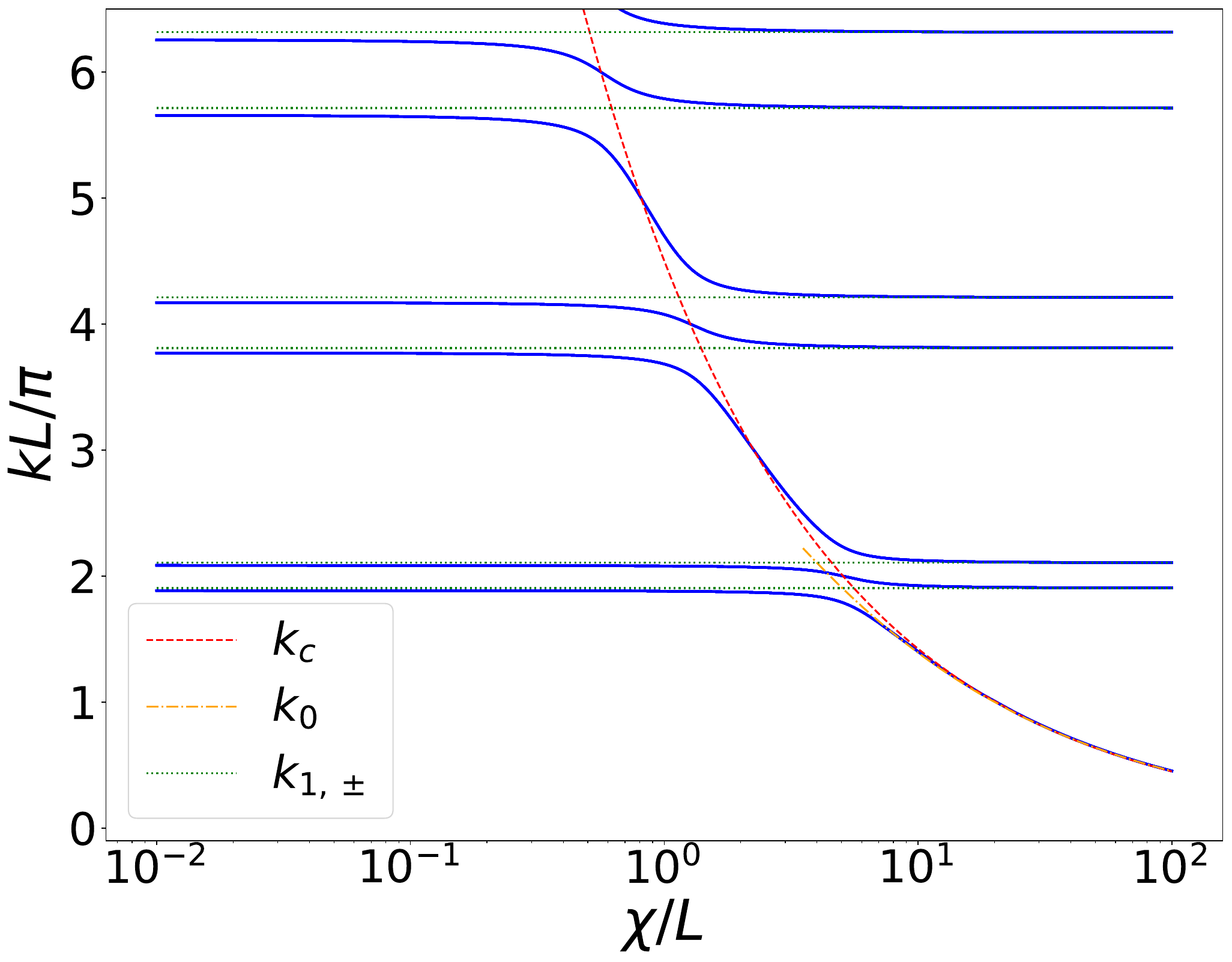}
         \caption{$\Delta L/L=0.05$, $vL=200$}
         \label{fig:ch2_generico_chi_critico}
     \end{subfigure}
     \begin{subfigure}[h]{0.49\columnwidth}
         
         \includegraphics[width=\linewidth]{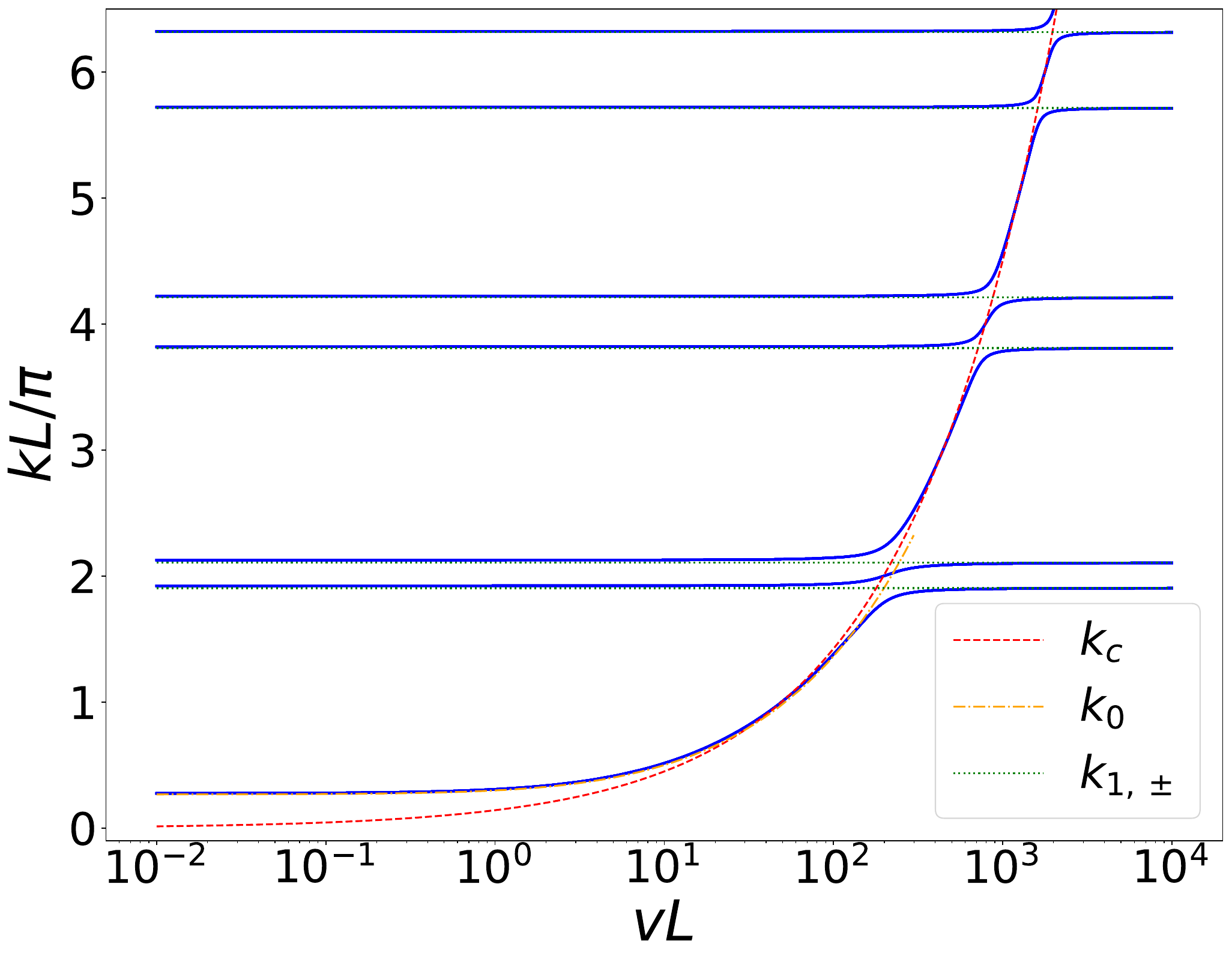}
         \caption{$\Delta L/L=0.05$, $\chi/L=5$}
         \label{fig:ch2_generico_v_critico}
     \end{subfigure}
    \caption{Eigenfrequencies dependence on each adimensional parameter of the cavity. All parameters are fixed except for the  asymmetry (a), the  susceptibility (b), and the  conductivity (c). The cavity considered for these plots is of unit length. The red curves represent the critical energy level $k_c$ as defined in Eq. \eqref{eq:k_critico}. Asymptotic ($k_{n,\pm}$) and low-energy  ($k_0$) solutions are also shown in green and orange, respectively.}
    \label{fig:cortes_planos}

\end{figure}
\end{widetext}

Eq. \eqref{eq:trascendente adimensional} defines a four-dimensional surface $a_1(a_2,a_3,a_4)$ from which we can then obtain the energy levels by computing $k=a_1/L$. To study the dependence of the energy levels on the cavity's parameters, FIG. \ref{fig:cortes_planos} shows plane cuts of the four-dimensional surface by fixing two out of three parameters in $\{a_i\}_{i=2,3,4}$. 

FIG. \ref{fig:ch2_generico_dL_critico} depicts a mixed behaviour: for $k<2\pi/L$ energy bands are similar to those of FIG. \ref{fig:ch2_standardDCE_variandov}, while for $k>2\pi/L$ spectrum resembles FIG. \ref{fig:ch2_standardDCE_variandochi}. However, a different behaviour is observed for $k\simeq 2\pi/L$, near the critical energy $k_c$. Although $k=k_c$ is typically not a solution of Eq. (\ref{eq:trascendente_estatica}), it clearly represents a change in the curvature of the energy bands. From now on, we will say the system is in a high-susceptibility regime if $k>k_c$ (if $\chi>0$) or in a high-conductivity regime if $k<k_c$. The latter may not exist even for non-zero values of $v$. For instance, in a high-susceptibility regime with $k_cL<\pi$ no low-energy modes are allowed and thus no solution with $k<k_c$ exists. As expected, FIG. \ref{fig:ch2_generico_dL_critico} is symmetrical with respect to the change of sign in $\Delta L$, as no physical differences exist between the left and right cavities. Again, energy bands do not overlap, but the gap can be reduced by tuning the properties of the dielectric membrane. For $k\gg k_c$, the gap is neglectable and the results are similar to those of the limit ($\chi\to\infty$). An analogous behaviour can be seen if instead the value of $v$ is dominant at the highest energies plotted.

On the other hand, FIG. \ref{fig:ch2_generico_chi_critico} and FIG. \ref{fig:ch2_generico_v_critico} show a splitting with avoided crossing in the eigenfrequencies of the system: where the energy bands are supposed to merge, now two bands are seen (see FIG. \ref{fig:merged}). For this to happen, the cavity must be asymmetrical and the dielectric membrane non-transparent (see Eq. \eqref{eq:soluciones_altasus}). As we increase the susceptibility (conductivity) in FIG. \ref{fig:ch2_generico_chi_critico} (FIG. \ref{fig:ch2_generico_v_critico}) energy levels are decreased (increased). For all energy bands, the most abrupt transition occurs at $k\simeq k_c$. Then Eq. \eqref{eq:k_critico} effectively predicts where these energy changes occur, as the system moves from a high-susceptibility regime to a high-conductivity regime (or vice versa). This is true for arbitrarily high or low energy levels. In FIG.  \ref{fig:ch2_generico_chi_critico}, a low-energy mode becomes available for $\chi/L\gtrsim10$. On the other hand, the initially allowed low-energy mode disappears in FIG. \ref{fig:ch2_generico_v_critico} for $vL\gtrsim 100$.

Asymptotic solutions can be found when the following condition is satisfied:
\begin{equation}
   |k^2\chi L-vL|\gg1, \label{eq:alta_sus}
\end{equation}
this occurs both in the high-susceptibility and high-conductivity regimes. If $kL\gtrsim1$ (untrue for the low-energy level), then Eq. \eqref{eq:trascendente_estatica} can be simplified using Eq. \eqref{eq:alta_sus}:
\begin{equation}
    \cos(k\Delta L)=\cos(kL),
\end{equation}
yielding the asymptotic solutions
\begin{equation}
    k_{n, \pm}=\dfrac{2n\pi}{L}\dfrac{1}{1\mp |\Delta L/L|}, ~n\in\mathbb{Z}^+. \label{eq:soluciones_altasus}
\end{equation}
Eq. \eqref{eq:soluciones_altasus} shows that the asymptotic degeneration of energy levels seen in FIG. \ref{fig:merged} can only happen for perfectly symmetrical cavities ($\Delta L=0$). Even for a slightly asymmetrical cavity (as the one in FIG. \ref{fig:cortes_planos}), this degeneracy is completely broken.

Additionally, we can obtain an approximate solution for the low-energy mode $k_0$ (only if in a high-susceptibility regime) by performing a fourth order Taylor expansion with respect to $k_0L$ in Eq. \eqref{eq:trascendente_estatica}:
\begin{equation}
    k_0=\frac{1}{L}\sqrt{\dfrac{4+vL\left(1-\frac{\Delta L^2}{L^2}\right)}{\frac{2}{3}+\frac{\chi}{L}\left(1-\frac{\Delta L^2}{L^2}\right)}}. \label{eq:modo_cero}
\end{equation}

If $\frac{\chi}{L}\left(1-\frac{\Delta L^2}{L^2}\right)\gg1$ and $vL\left(1-\frac{\Delta L^2}{L^2}\right)\gg1$, then Eq. \eqref{eq:modo_cero} simplifies to the critical energy level $k_c$ from Eq. \eqref{eq:k_critico}. It is for this reason that the critical curve accurately describes $k_0$ in FIG. \ref{fig:ch2_generico_chi_critico}, but not in FIG. \ref{fig:ch2_generico_v_critico} (as in some regions of FIG. \ref{fig:ch2_generico_v_critico} the condition $vL\gg1$ is not satisfied). The asymptotic and low-energy level approximations are also shown in FIG. \ref{fig:cortes_planos}.

FIG. \ref{fig:cortes_planos} shows a great agreement between the solutions of Eq. \eqref{eq:trascendente adimensional} and the asymptotic/low-energy solutions of Eq. \eqref{eq:soluciones_altasus} and Eq. \eqref{eq:modo_cero}. However, on the left side of FIG. \ref{fig:ch2_generico_chi_critico} numeric solutions do not exactly match the asymptotic ones. This is due to a low conductivity not satisfying Eq. \eqref{eq:alta_sus}. On the other hand, in FIG. \ref{fig:ch2_generico_v_critico} the conductivity is high enough on the right side of the plot, while the susceptibility is high enough on the left side. Additionally, the low-energy mode approximation (Eq. \eqref{eq:modo_cero}) is valid even beyond $k_0L\ll1$. It only starts to fail once the high-susceptibility regime is shifting towards a high-conductivity one (at least at low energies), and modes with such a low energy are prohibited.

\subsection{Complex solutions ($\dot{\chi}\neq0$)}

\begin{figure}[H]
     \centering
     \begin{subfigure}[h]{0.9\columnwidth}
         \centering
         \includegraphics[width=\textwidth]{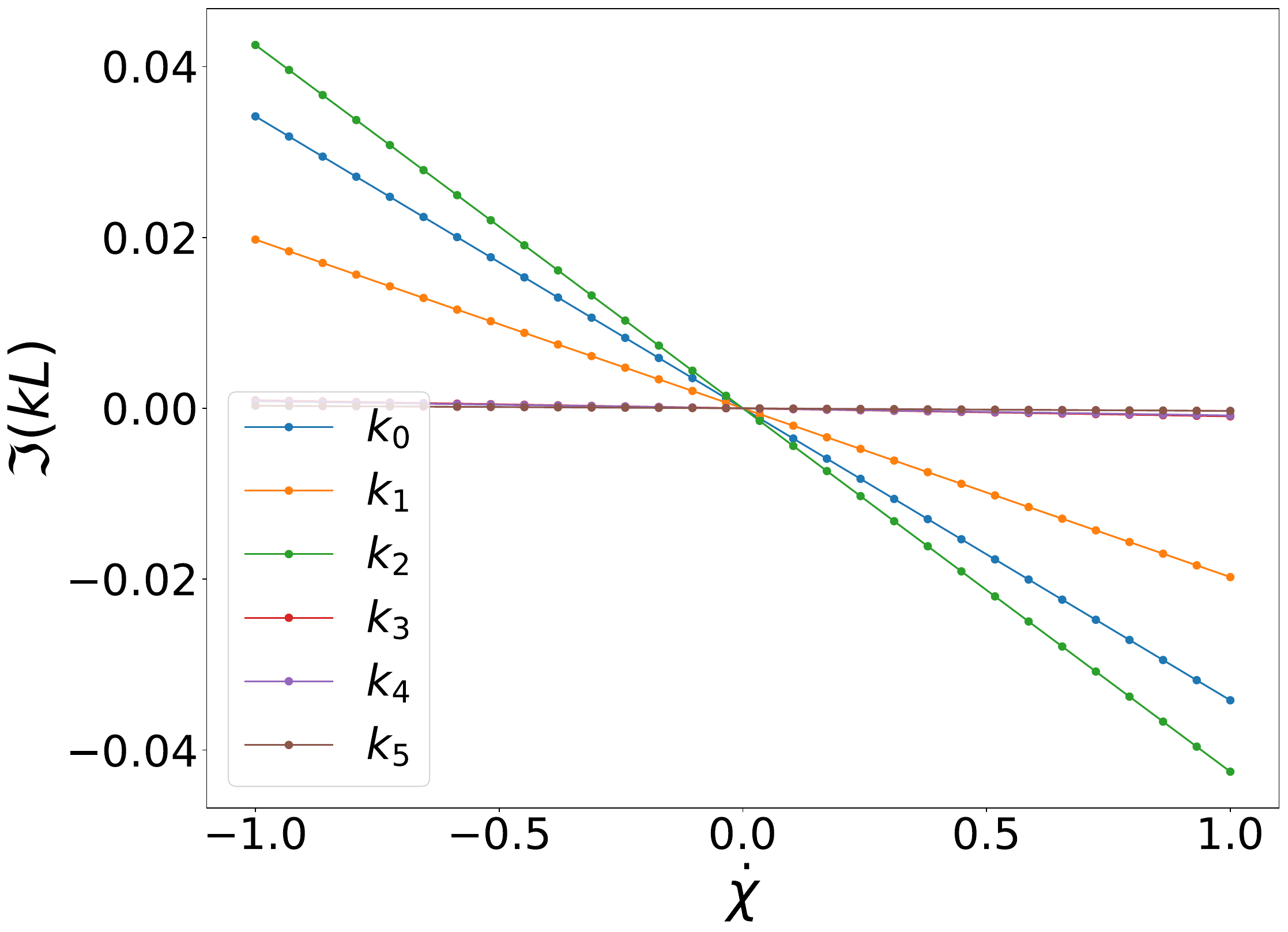}
         \caption{}
         \label{fig:ch2_parteim_chipunto}
     \end{subfigure}
     \\
     \begin{subfigure}[h]{0.9\columnwidth}
         \centering
         \includegraphics[width=\textwidth]{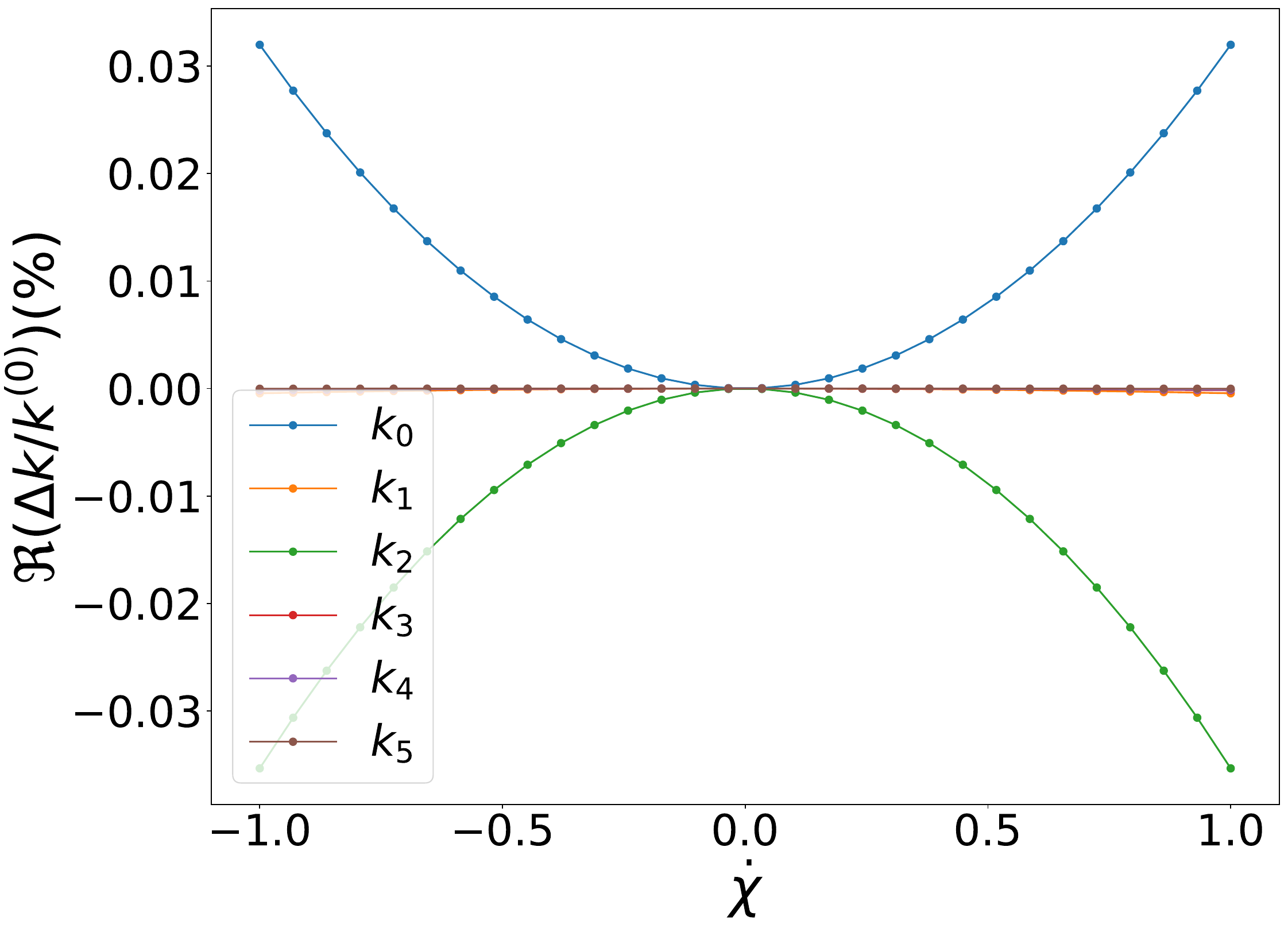}
         \caption{}
         \label{fig:ch2_partere_chipunto}
     \end{subfigure}
        \caption{ Complex solutions of Eq. \eqref{eq:trascendente_estatica} as a function of $\dot{\chi}$, for a cavity with $\Delta L/L=0.05$, $\chi/L=5$, $vL=200$ and unit total length. Imaginary part of the solutions is shown in (a), while the percentage change in the real part is shown in (b). $k^{(0)}$ is the solution for $\dot{\chi}=0$.}
        \label{fig:ch2_partes_chipunto}
\end{figure}

As previously mentioned, the instantaneous solutions of Eq. \eqref{eq:trascendente_estatica} acquire a non-zero imaginary part when the cavity is not static and the susceptibility is being tuned (i.e. $\dot{\chi}\neq0$). 
FIG. \ref{fig:ch2_partes_chipunto} shows how the instantaneous solutions change due to having $\dot{\chi}\neq0$, up to a 20\% of $\chi/L$. The imaginary part is directly proportional to $\dot{\chi}$, for all considered values of $\dot{\chi}$. The imaginary part is almost neglectable with respect to the static solution, even for the lowest energy level. For higher energy levels, the imaginary part can't even be appreciated in the plot. On the other hand, the change in the real part is parabolic and completely neglectable, as seen in FIG. \ref{fig:ch2_partere_chipunto}. In both cases, the most affected energy levels are the lowest one (blue) and the closest one to the critical energy level (green).

\section{Eigenfunctions localization} \label{sec:eigenfunc}

\begin{figure}[H]
    \centering
    \includegraphics[width=0.99\columnwidth]{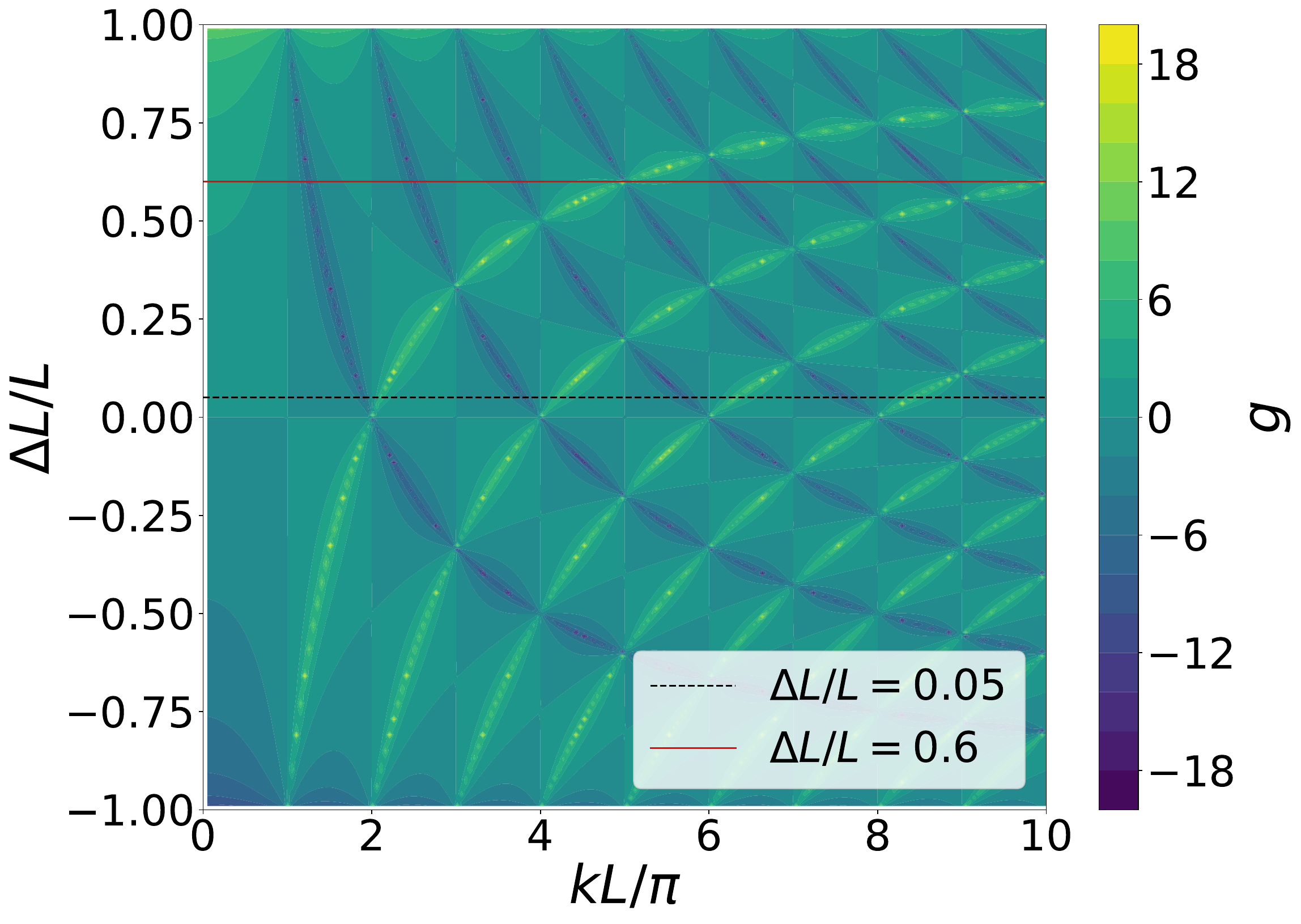}
    \caption{Localization degree for a generic cavity as a function of the normalized eigenfrequency and asymmetry. For a slightly asymmetric double cavity (black dotted line), we see that low energy modes might alternate between being localized at one cavity or the other. For a very asymmetric one (red continuous line), it will be much easier to localize modes in the larger cavity. In the plot, this is seen as the red line crosses many purple regions before it crosses a yellow one.}
    \label{fig:ch2_kappa_contour}
\end{figure}

\begin{figure}[H]
     \centering
         \includegraphics[width=.49\textwidth]{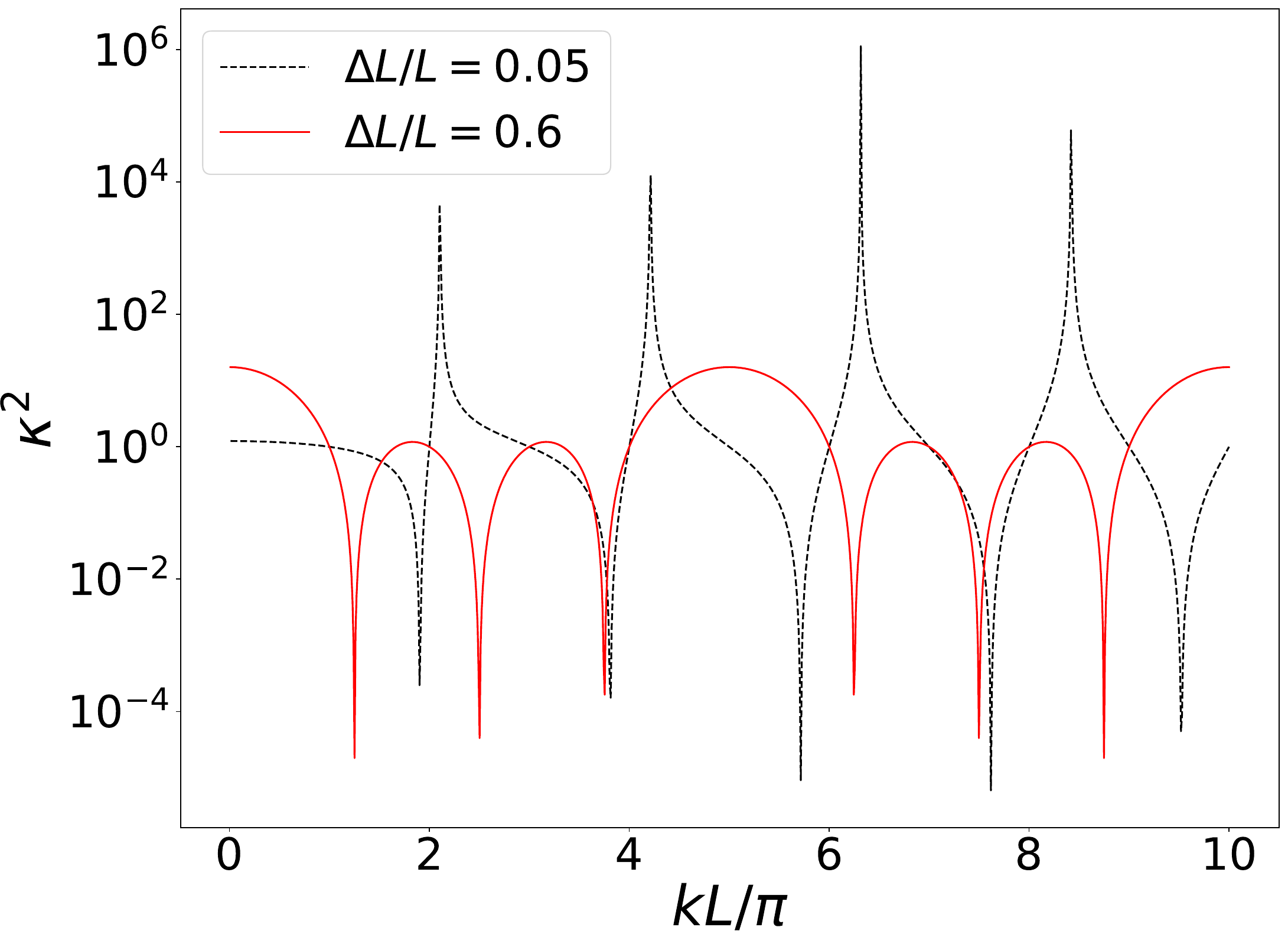}
        \caption{Localization constant $\kappa^2$ as a function of the normalized eigenfrequencies for a slightly asymmetric cavity with $\Delta L/L=0.05$ (black dotted line) and a very asymmetric one with $\Delta L/L=0.6$   (red line). In both cases, the double cavity is of unit length and $\chi$, $v$ are free parameters. These curves are cuts of the plot in in FIG. \ref{fig:ch2_kappa_contour} across lines of fixed $\Delta L$ values.}
        \label{fig:kappa_line}
\end{figure}

Following the solution basis defined in Eq. \eqref{eq:modos_espaciales_dinamicos}, it is clear that the constant multiplying the right cavity wavefunction allows us to have localized and delocalized eigenfunctions. We define $\kappa$ as the absolute value of this constant:
\begin{equation}
    \kappa^2\equiv\left|\frac{\sin(kL_1)}{\sin(kL_2)}\right|^2. \label{eq:localizacion}
\end{equation}
If $\kappa^2\sim 1$, then the eigenfunction is (mostly) delocalized. If instead $\kappa^2\gg1$ ($\kappa^2\ll1$), then the eigenfunction is strongly localized in the right (left) cavity. In FIG. \ref{fig:ch2_kappa_contour} we show how localized modes can be for generic cavities. For slightly asymmetric cavities (black dotted line in FIG. \ref{fig:ch2_kappa_contour}, see also FIG. \ref{fig:kappa_line}), we see that a strong localization can be seen in both cavities. Also, only a small energy difference is enough to completely localize modes in different cavities. Given the similar length of the cavities, the modes alternate between being localized at the right and left cavities. This alternation between consecutive modes does not hold for higher energy levels. On the other hand, for very asymmetric cavities (red line in FIG. \ref{fig:ch2_kappa_contour} and FIG. \ref{fig:kappa_line}), localization is much easier in the longest cavity (left) at low energy levels. This is why we see a similar localization degree in the left cavity, while it is impossible to achieve $\kappa^2\gg1$ and localize a mode in the right cavity. Of course, increasing the energy allows for localization in the right cavity.

We can choose to have specific modes localized in any cavity, or delocalized. To achieve this, we must first note that $\kappa^2(k)$ is only dependent on two additional parameters as per Eq. \eqref{eq:localizacion}: $L$ and $\Delta L$. The first step then consists in adequately choosing $L$ and $\Delta L$ so that a specific mode $k_p$ satisfies $\kappa^2(k_p)=\kappa_p^2$, where $\kappa_p^2$ is the desired localization degree. The second and last step is to ensure that $k_p^2$ is a solution of Eq. \eqref{eq:trascendente_estatica}, for which we can choose the electric properties of the membrane: $\chi$ and $v$. 

As we are interested in knowing if $\kappa^2$ is much bigger than, much smaller than or similar to $1$, and as $\kappa^2>0$, $g=2\ln\kappa$ defines in a more direct way the localization of the eigenfunctions. For $g\sim0$ then the mode is delocalized, while for $|g|>0$ the mode is localized in the right ($g>0$) or left ($g<0$) cavity. FIG. \ref{fig:ch2_kappa_contour} shows the values of $g$ for generic cavities, as a function of the normalized eigenfrequency and asymmetry of the cavity. As expected, $g$ is anti-symmetric with respect to the cavity exchange: $g(k,\Delta L)=-g(k,-\Delta L)$. It also shows that eigenfunctions are typically delocalized with the exception of some that lie inside the narrow yellow or purple regions in FIG. \ref{fig:ch2_kappa_contour}. The curves that describe this regions are those that  make the denominator or numerator in Eq. \eqref{eq:localizacion} go to zero: $k_n=n\pi/L_1$ (purple) and $k_n=n\pi/L_2$ (yellow). For sufficiently high values of $|\Delta L|$, modes stop localizing at one of the cavities until the energy is increased enough. This is consistent with the previous discussion on FIG. \ref{fig:kappa_line}. Lastly, modes with energies lower than $\pi/L$ cannot be strongly localized for any value of $\Delta L$.

\section{Photon creation}\label{sec: photon}

The dynamics of the $Q_n^{(m)}(t)$ coefficients of the field modes (Eq. \eqref{eq:ecuacion_Q}) have no analytical solutions. Thus, it must be studied under numeric calculations and analytical approximations. To approach our problem, we consider the Multiple Scale Analysis (MSA) approximation \cite{velasco2022photon, crocce2002quantum, lombardo2016dynamical, setare2008particle, yuce2008dynamical} which consists on defining a second, slower time scale. The MSA approximation yields solutions valid at longer times than those of perturbative methods, by avoiding secular terms. In the next sections, we present different solutions obtainable under different coupling conditions required by the  MSA approximation. A complete deduction can be found in Appendix \ref{ap:MSA}.

For the MSA method we consider a very restricted subset of possible trajectories of the cavity parameters: sinusoidal perturbations with small amplitudes. Additionally, we ask that all amplitudes be of the same order of magnitude. This condition allows us to consider a single additional time scale.

As shown in Appendix \ref{ap:MSA}, we can get non trivial solutions for only specific couplings between the driving frequencies and the spectrum of the cavity. For at least one parameter $r$, at least one of the following conditions must be met: 
\begin{align}
\Omega&=2\omega_l \label{eq:res_param},\\
\Omega&=\omega_l+\omega_n \label{eq:suma},\\
\Omega&=\omega_l-\omega_n\label{eq:resta1},\\
\Omega&=\omega_n-\omega_l \label{eq:resta2},
\end{align}
where $\Omega_r$ is the external driving frequency of the cavity parameter $r$.
We will refer to the condition imposed by Eq. \eqref{eq:res_param} as parametric resonance. The remaining equations (Eq. \eqref{eq:suma} to Eq. \eqref{eq:resta2}) present a coupling between two distinct cavity modes and can be simply written as:
\begin{equation}
    \Omega=|\omega_n\pm\omega_l|.
\end{equation}
For the $+$ symbol and $n=l$, this expression contains the  parametric resonance.

\subsection{Parametric resonance}
Parametric resonance is of special interest because it is known to be possible to obtain an exponentially-fast growing number of photons by adequately choosing the driving of a mode in a non equidistant region of the cavity spectrum \cite{Dodonov96, crocce2001resonant, dalvit1999creation}. 

We define the following coefficients that relate to resonant coupling ($\eta$) and two-mode coupling ($\beta$, $\alpha$):
\begin{align}
    \eta_n^{(r)}&\equiv\frac{\partial \omega_n}{\partial_r},\label{eq:eta_definicion}\\
    \beta_{nl}^{(r)}&\equiv (\partial_r \varPhi_n, \varPhi_l), \label{eq:beta} \\
    \alpha_{nl}^{(r)}&\equiv (\partial_r^2 \varPhi_n, \varPhi_l).\label{eq:alfa}
\end{align}
All the above derivatives are taken at $t=0$.

Let's assume we have a cavity with a non-equidistant spectrum such that it exists $\omega_l$ that satisfies $2\omega_l\neq \omega_m\pm\omega_n$, $\forall (n,m)\neq (l,l)$. For simplicity, let's consider equal external driving frequencies on all parameters $\Omega=2\omega_l$. Because $\beta_{nn}^{(r)}=0$ except for $\chi$ and $\dot{\chi}$ (Eq. \eqref{eq:apC_betagenerico}), and using Eq. \eqref{eq:MSA_A}, Eq. \eqref{eq:MSA_B}, Eq. \eqref{eq:apB_cociente1} and Eq. \eqref{eq:apC_relacionbetachichipunto}:
\begin{align}
\frac{dA_l^m}{d\tau}&=B_l^m\left\{-\xi_\chi\chi_0 
\eta_l^{(\chi)} 
+\frac{1}{2}\sum_{r\neq \dot{\chi}} \xi_rr_0
 \eta_l^{(r)}\right\},
 \label{eq:ch3_res_param_A}\\
\frac{dB_l^m}{d\tau}&=A_l^m\left\{ +\xi_\chi\chi_0\eta_l^{(\chi)}
+\frac{1}{2}\sum_{r\neq \dot{\chi}} \xi_rr_0 \eta_l^{(r)}  \right\}.
\label{eq:ch3_res_param_B}
\end{align}
Differentiating Eq. \eqref{eq:ch3_res_param_B} again with respect to  $\tau$ and by using Eq. \eqref{eq:ch3_res_param_A} we obtain:
\begin{align}
\frac{d^2B_l^m}{d\tau^2}&=B_l^m\left\{\Gamma^2-\Lambda^2\right\},\\
\Gamma&\equiv\frac{1}{2}\sum_{r\neq \dot{\chi}} \xi_rr_0 \eta_l^{(r)},\label{eq:def_Gamma}\\
\Lambda&\equiv\xi_\chi\chi_0\eta_l^{(\chi)}.
\end{align}

\begin{figure}[H]
    \centering
    \includegraphics[width=0.9\columnwidth]{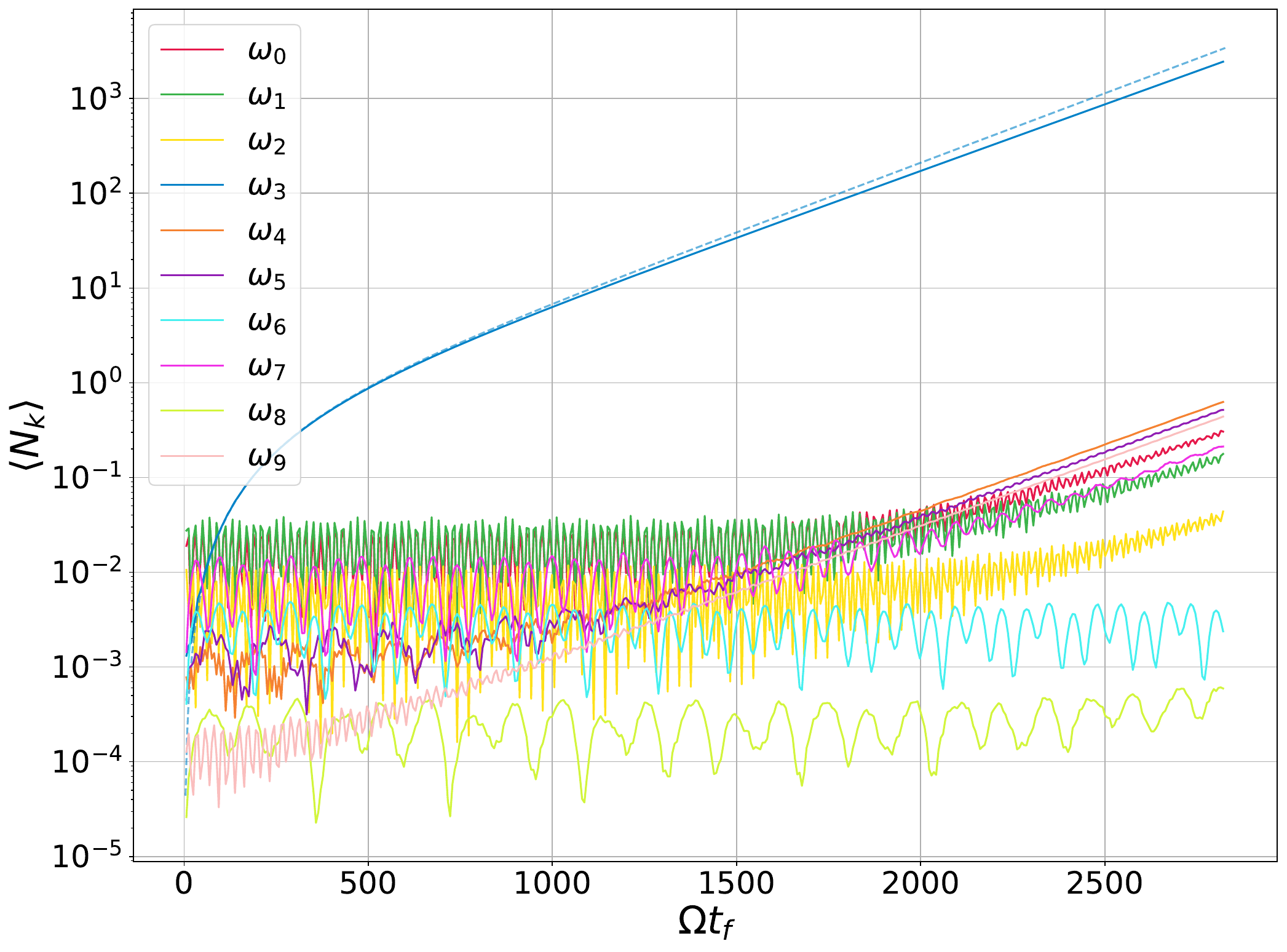}
    \caption{Expected photon number for each cavity mode as a function of time. The dotted line corresponds to a \textit{Multiple Scale Analysis} solution. The system was trimmed at the first $10$ modes, with a cavity with $\Delta L_0/L_0=0.44$, $\chi_0/L_0=0.5$, $v_0L_0=0$ and of unit length. The asymmetry $\Delta L$ was tuned with $\epsilon=0.01$ and $\Omega=2\omega_3$.} 
    \label{fig:ch4_res_param_-1}
\end{figure}

Let $\gamma^2\equiv|\Gamma^2-\Lambda^2|$. Because $\Gamma$ and $\Lambda$ are strictly real, we expect three different behaviours for $A$ and $B$. If $|\Gamma|>|\Lambda|$, the solution will be a linear combination of real exponential functions, and the photon number will grow exponentially at a rate proportional to $\epsilon\gamma$. Considering the initial conditions imposed by Eq. \eqref{eq:ch3_cond_inicial_A} and Eq, \eqref{eq:ch3_cond_inicial_B}, the most general solution is given by:
\begin{align}
    A_l^{(m)}(\tau) &=\frac{\delta_{lm}}{\sqrt{2\omega_l}}\frac{\gamma}{\Gamma+\Lambda}\sinh(\gamma \tau),\\
    B_l^{(m)}(\tau)
    &=\frac{\delta_{lm}}{\sqrt{2\omega_l}}\cosh(\gamma\tau),
\end{align}
and with Eq. \eqref{eq:numero_fotones} we can obtain the expected number of photons in the resonant mode:
\begin{equation}
    \left\langle N_l\right\rangle(t)=\frac{\Gamma-\Lambda}{\Gamma+\Lambda}\left(1+N_l^{(0)}\right)\sinh^2(\epsilon\gamma t)+N_l^{(0)}\cosh^2(\epsilon\gamma t), \label{eq:fotones_res_param_exp}
\end{equation}
where we see that the photon production rate is  $2\epsilon\gamma$. Photon number of every other frequency is expected to remain constant. However, this doesn't hold true for long times as shown in FIG. \ref{fig:ch4_res_param_-1}. We see that the MSA approximation (dotted blue line) predicts the behaviour of the system at times $\Omega t_f\lesssim 2000$, where $\Omega$ is the chosen driving frequency. However, at longer times, MSA predicts higher photon generation rates than the real solution does and fails to predict an exponential surge in other cavity modes. The exponential growths in other modes start at different times, and eventually reaches the same rate as the resonant mode.

\begin{figure}[H]
    \centering
    \includegraphics[width=0.9\columnwidth]{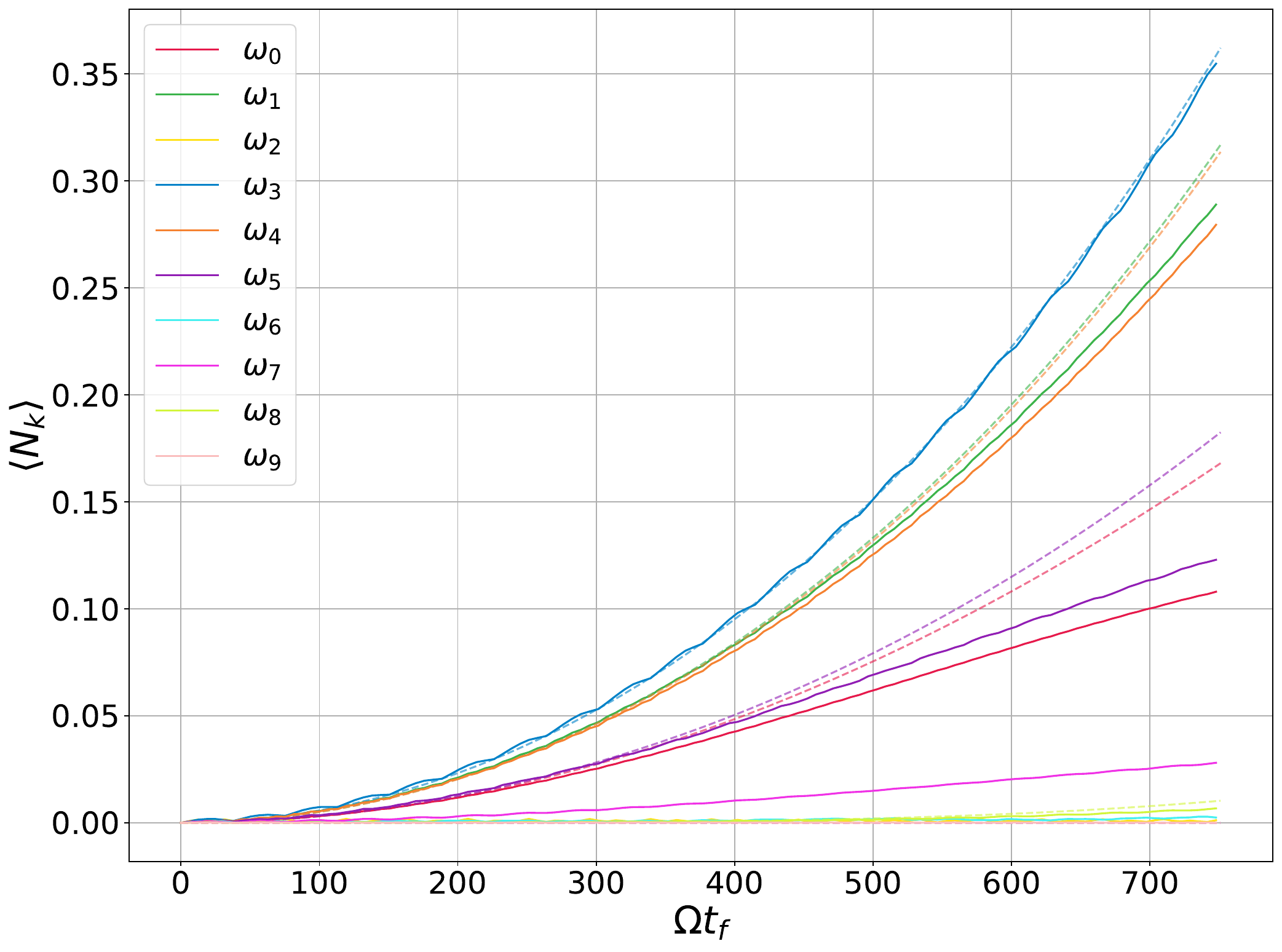}
    \caption{Expected photon number for each cavity mode as a function of time. The dotted lines correspond to a \textit{Multiple Scale Analysis} solutions allowing a certain non-zero detuning from the driving frequency. The system was trimmed at the first $10$ modes, with a cavity with $\Delta L_0/L_0=0.44$, $\chi_0/L_0=0.5$, $v_0L_0=200$ and of unit length. The asymmetry $\Delta L$ was tuned with $\epsilon=0.01$ and $\Omega=2\omega_3$.}
    \label{fig:ch4_res_param_2}
\end{figure}

FIG. \ref{fig:ch4_res_param_2} shows the importance of the election of the driving frequency and the cavity parameters. If a mode with weaker self-coupling coefficients $\eta$ is driven, then the generated photons will be much less. Also, even for a non equidistant cavity spectrum, we see that other modes can couple and display exponential growth. In the particular situation of FIG. \ref{fig:ch4_res_param_2}, there are two couplings that are slightly ($<0.5$\%) detuned from the driving frequency:
\begin{align}
    \frac{\omega_1+\omega_4}{2\omega_3}-1&=0.00088
    ,\\\frac{\omega_0+\omega_5}{2\omega_3}-1&=-0.0038.
\end{align}

The other dotted lines in FIG. \ref{fig:ch4_res_param_2} correspond to the MSA solutions if the Kronecker delta functions are replaced with step functions with a small width. This simple relaxation of the delta functions allows us to predict how photons are produced in other mode, and even accurately predicts which of the coupled modes will exhibit a higher number of photons (green and purple, and not orange or red). At longer times  these MSA solutions overestimate the number of photons in every mode.

\begin{figure}[H]
    \centering
    \includegraphics[width=0.9\columnwidth]{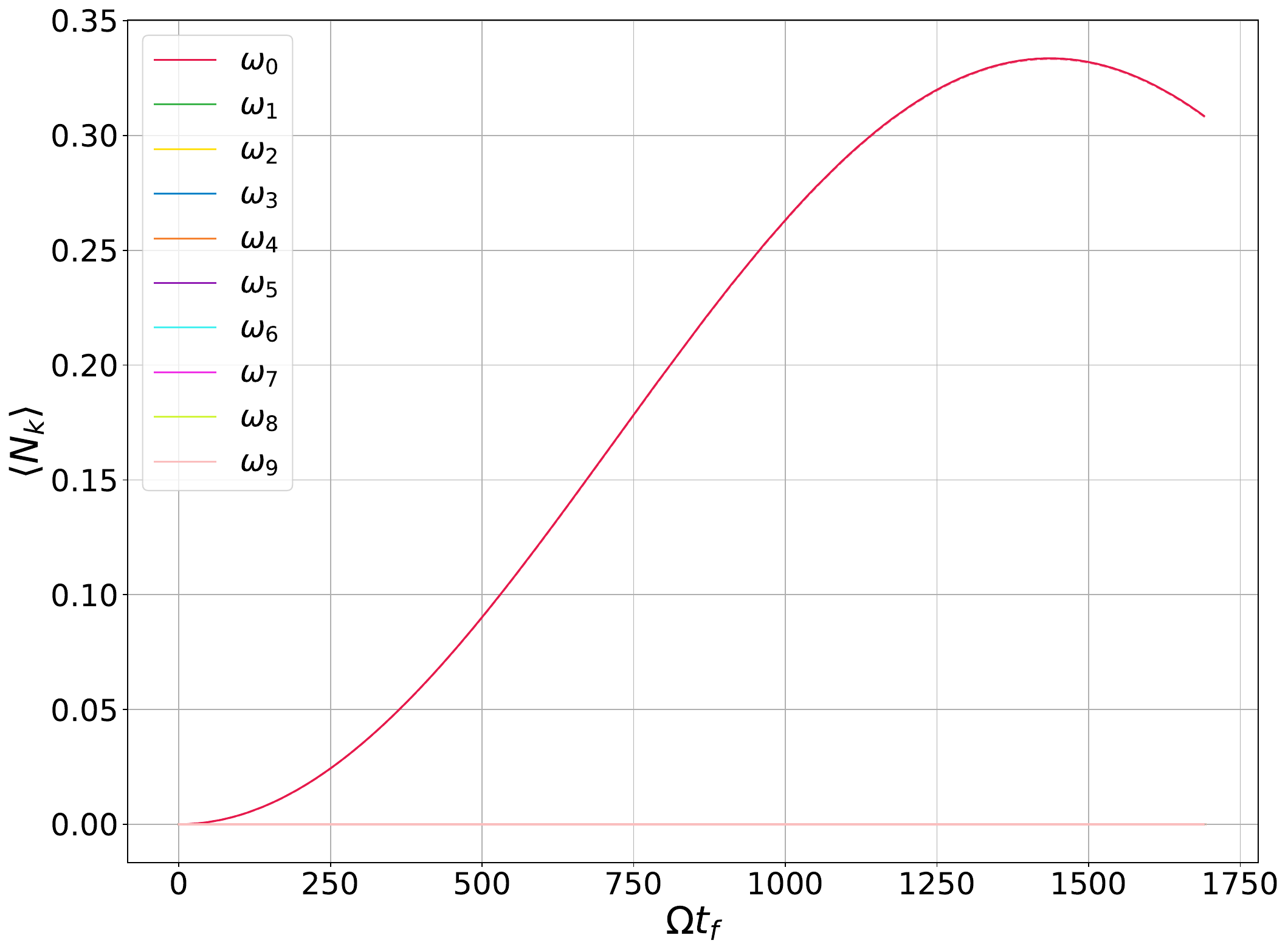}
    \caption{Expected photon number for each cavity mode as a function of time. The dotted line corresponds to a \textit{Multiple Scale Analysis} solution. The system was trimmed at the first $10$ modes, with a cavity with $\Delta L_0/L_0=0.44$, $\chi_0/L_0=0.5$, $v_0L_0=0$ and of unit length. The susceptibility $\chi$ was tuned with $\epsilon=0.01$ and $\Omega=2\omega_0$.} 
    \label{fig:ch4_res_param_chi}
\end{figure}

If instead $|\Lambda|>|\Gamma|$, then the photon number will oscillate at a frequency proportional to $\epsilon\gamma$, as shown in FIG. \ref{fig:ch4_res_param_chi}:
\begin{align}
    A_l^{(m)}(\tau)&=-\delta_{lm}\frac{1}{\sqrt{2\omega_l}}\frac{\gamma}{\Gamma+\Lambda}\sin(\gamma \tau),\\
    B_l^{(m)}(\tau)&=\frac{\delta_{lm}}{\sqrt{2\omega_l}}\cos(\gamma\tau),
\end{align}

\begin{equation}
    \left\langle N_l\right\rangle(t)=\frac{\Lambda-\Gamma}{\Gamma+\Lambda}\left(1+N_l^{(0)}\right)\sin^2(\epsilon\gamma t)+N_l^{(0)}\cos^2(\epsilon\gamma t)\label{eq:res_param_generica}.
\end{equation}

 This oscillations arise even when the cavity is prepared in a vacuum state. This scenario where $|\Lambda|>|\Gamma|$ can be seen as a shortcut to adiabaticity as all photons produced by the tuning of the parameters are eventually destroyed after a certain finite time, if the cavity remains unobserved. However, if there are $N_l^{(0)}$ photons in the \textit{in} region, then this procedure will never get the expected photon number in the cavity below $N_l^{(0)}$. Therefore, it can be used to change the cavity parameters without producing photons but cannot be used to cool the cavity down. The number of photons that can be generated (amplitude of the oscillations) is proportional (disregarding a constant) to $\frac{1-\alpha}{1+\alpha}$, with $\alpha=\Gamma/\Lambda$, $|\alpha|<1$. If $\Gamma<0$, by increasing $|\Gamma|$ we can achieve higher amplitudes.

The last case is the least relevant one, as it would be impossible to observe in the lab ($\gamma$ must be exactly zero). If $\Gamma=\Lambda$, coefficients remain constant over time. If $\Gamma=-\Lambda$, then:
\begin{align}
    A_l^{(m)}(\tau)&=-\delta_{lm}\frac{2\Lambda}{\sqrt{2\omega_l}}\tau,\\
    B_l^{(m)}(\tau)&=\frac{\delta_{lm}}{\sqrt{2\omega_l}},
\end{align}
and photon number grows quadratically in time:
\begin{equation}
    \left\langle N_l\right\rangle(t)=4\left(1+N_l^{(0)}\right)\Lambda^2\epsilon^2t^2+N_l^{(0)}.
\end{equation}

It must be noted that this behaviour is consistent with previous cases, and the transition implies no discontinuity. If $\Gamma\to-\Lambda$ in Eq. \eqref{eq:fotones_res_param_exp} and Eq. \eqref{eq:res_param_generica}, then $\gamma\to0$ and thus we obtain an infinite number of photons in an infinite time (and thus the parabola).

By allowing to tune multiple parameters of the cavity at the same time we could control the exponential generation rate, or even frustrate completely the typical exponential production. The latter is harder to achieve, as it requires a precise election of the driving amplitudes. For instance, let's consider the conditions of FIG. \ref{fig:ch4_res_param_-1}. Eq. \eqref{eq:def_Gamma} is a recipe to compute the right amplitude of a second driving. By making $\Gamma=0$, we can obtain:
\begin{equation}
    \xi_{L}=-\xi_{\Delta L}\frac{\Delta L_0 \eta_3^{(\Delta L)}}{L_0\eta_3^{(L)}}\simeq-0.43856,
\end{equation}
which completely avoids the exponential growth in the photon number of the resonant mode. It is important to note that, if the above value of $\xi_{L}$ wouldn't have been of order $\mathcal{O}(1)$, then the MSA approximation might not have accurately predicted the number of photons produced.

\begin{figure}[H]
    \centering
    \includegraphics[width=0.9 \columnwidth]{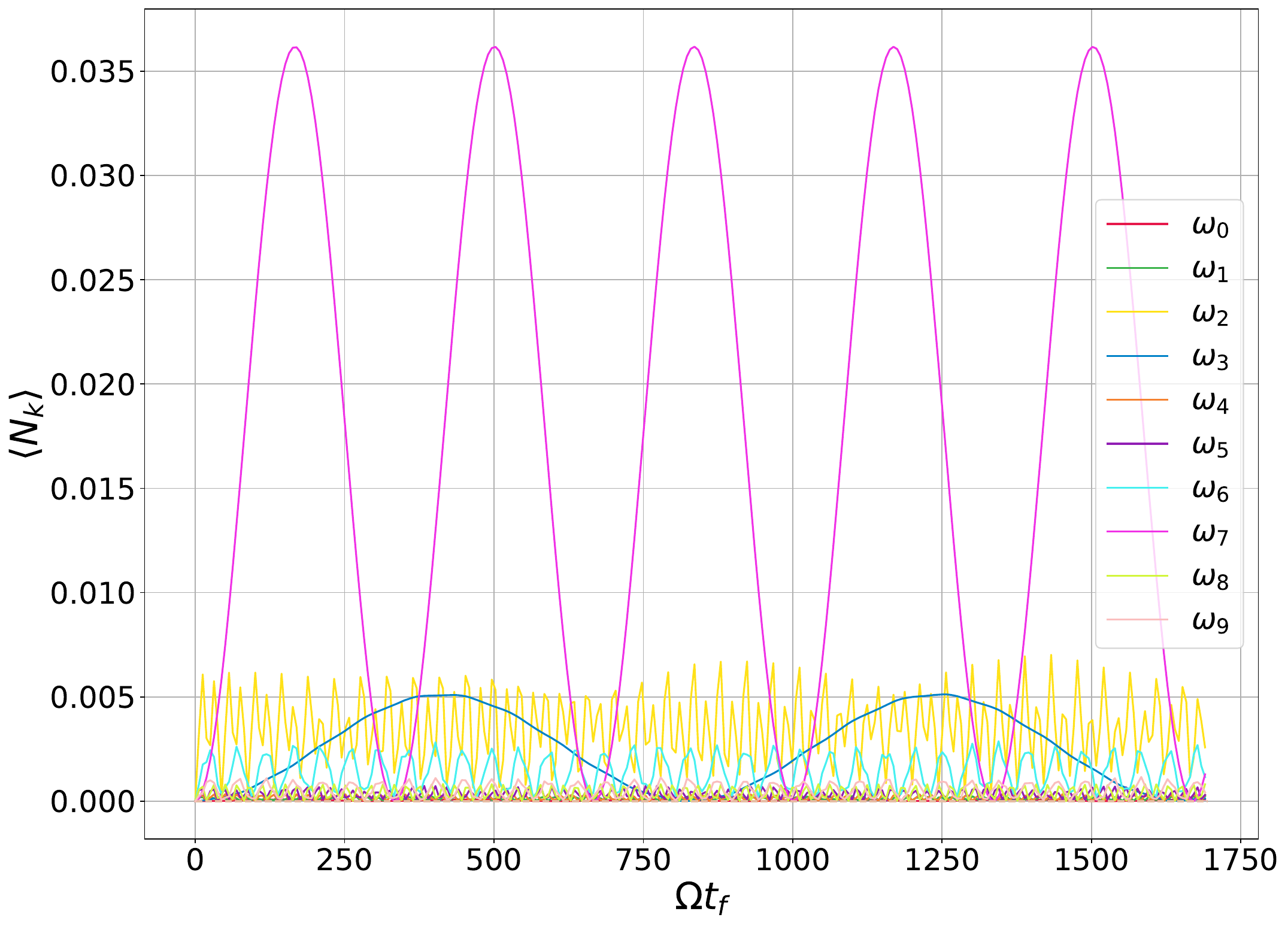}
    \caption{Complete frustration of parametric resonance under the conditions of FIG. \ref{fig:ch4_res_param_-1}, by considering a second tuning in the $L$ parameter with the same frequency but with  $\xi_L\simeq-0.43856$.} 
    \label{fig:ch4_res_param_LdL_cancelada}
\end{figure}

\subsection{Sum coupling}

Similar phenomena arise when coupling (additively) two cavity modes. It can be shown that if the driving frequency is $\Omega=\omega_n+\omega_l$, then the MSA solutions yield, for $\xi_\chi=0$:
\begin{align}
    \left\langle N_l\right\rangle(t)&=\left(1+N_n^{(0)}\right)\sinh^2(\epsilon\gamma t)+N_l^{(0)}\cosh^2(\epsilon\gamma t),\\
    \left\langle N_n\right\rangle(t)&=\left(1+N_l^{(0)}\right)\sinh^2(\epsilon\gamma t)+N_n^{(0)}\cosh^2(\epsilon\gamma t),\\
    \left\langle N\right\rangle(t)&=\left(2+N^{(0)}\right)\sinh^2(\epsilon\gamma t)+N^{(0)}\cosh^2(\epsilon\gamma t),
\end{align}
where
\begin{align}
\gamma&\equiv\frac{\omega_n^2-\omega_l^2}{4\sqrt{\omega_n\omega_l}}  \sum_{r\neq \chi,\dot{\chi}} r_0\xi_r\beta_{nl}^{(r)}. 
\end{align}
We now see exponential growth in both coupled modes, with an equal rate. This can happen even if the initial state is one with no photons.

A very different behaviour arises when we consider $\xi_\chi\neq0$ and (for simplicity) $\xi_r=0$ for all other parameters. We start by defining:
\begin{align}
    \tilde{\gamma}&\equiv\frac{\omega_n^2-\omega_l^2}{4\sqrt{\omega_n\omega_l}}\chi_0\xi_\chi,\\
    \gamma_{nl}&\equiv\tilde{\gamma} \sqrt{\left[1+\frac{2\omega_l}{\omega_n}\right]},
\end{align}
which allows us to write:
\begin{align}
\frac{d^2B_n^m}{d\tau^2}&=\gamma_{nl}^2 {\beta_{nl}^{(\chi)}}\beta_{ln}^{(\chi)}B_n^m,\\
\frac{dB_n^m}{d\tau}&=-\gamma_{nl} \beta_{ln}^{(\chi)} \sqrt{2+\frac{\omega_n}{\omega_l}}A_l^m.
\end{align}
$\gamma$ is a non-zero real number and according to Eq. \eqref{eq:apC_betagenerico}, $\beta_{nl}^{(\chi)}\beta_{ln}^{(\chi)}<0$. This implies that the number of photons will oscillate in time. We continue by defining:
\begin{align}
    \kappa_{nl}&\equiv\gamma_{nl}\sqrt{-\beta_{nl}^{(\chi)}\beta_{ln}^{(\chi)}}=\gamma_{nl}\left|\beta_{nl}^{(\chi)}\right|\frac{\omega_l}{\omega_n},\\
    \theta_{nl}&\equiv\sqrt{-\frac{\beta_{nl}^{(\chi)}}{\beta_{ln}^{(\chi)}}}\sqrt{1+2\frac{\omega_l}{\omega_n}}\frac{\omega_l}{\omega_n}=\sqrt{1+2\frac{\omega_l}{\omega_n}},
\end{align}
so now the solutions are:
\begin{align}
    A_l^{(m)}(\tau)&=-\frac{\delta_{nm}}{\sqrt{2\omega_l}}\theta_{nl}\sin(\kappa_{nl}\tau)\\
    B_n^{(m)}(\tau)&=\frac{\delta_{nm}}{\sqrt{2\omega_n}}\cos(\kappa_{nl}\tau).
\end{align}
Then we can finally write:
\begin{align}
    \left\langle N_l\right\rangle(t)&=\left(1+N_n^{(0)}\right)\theta_{nl}^2\sin^2(\epsilon\kappa_{nl} t)+N_l^{(0)}\cos^2(\epsilon\kappa_{ln} t),\\
    \left\langle N_n\right\rangle(t)&=\left(1+N_l^{(0)}\right)\theta_{ln}^2\sin^2(\epsilon\kappa_{ln} t)+N_n^{(0)}\cos^2(\epsilon\kappa_{nl} t).
\end{align}

\begin{figure}[H]
         \centering
         \includegraphics[width=.9\columnwidth]{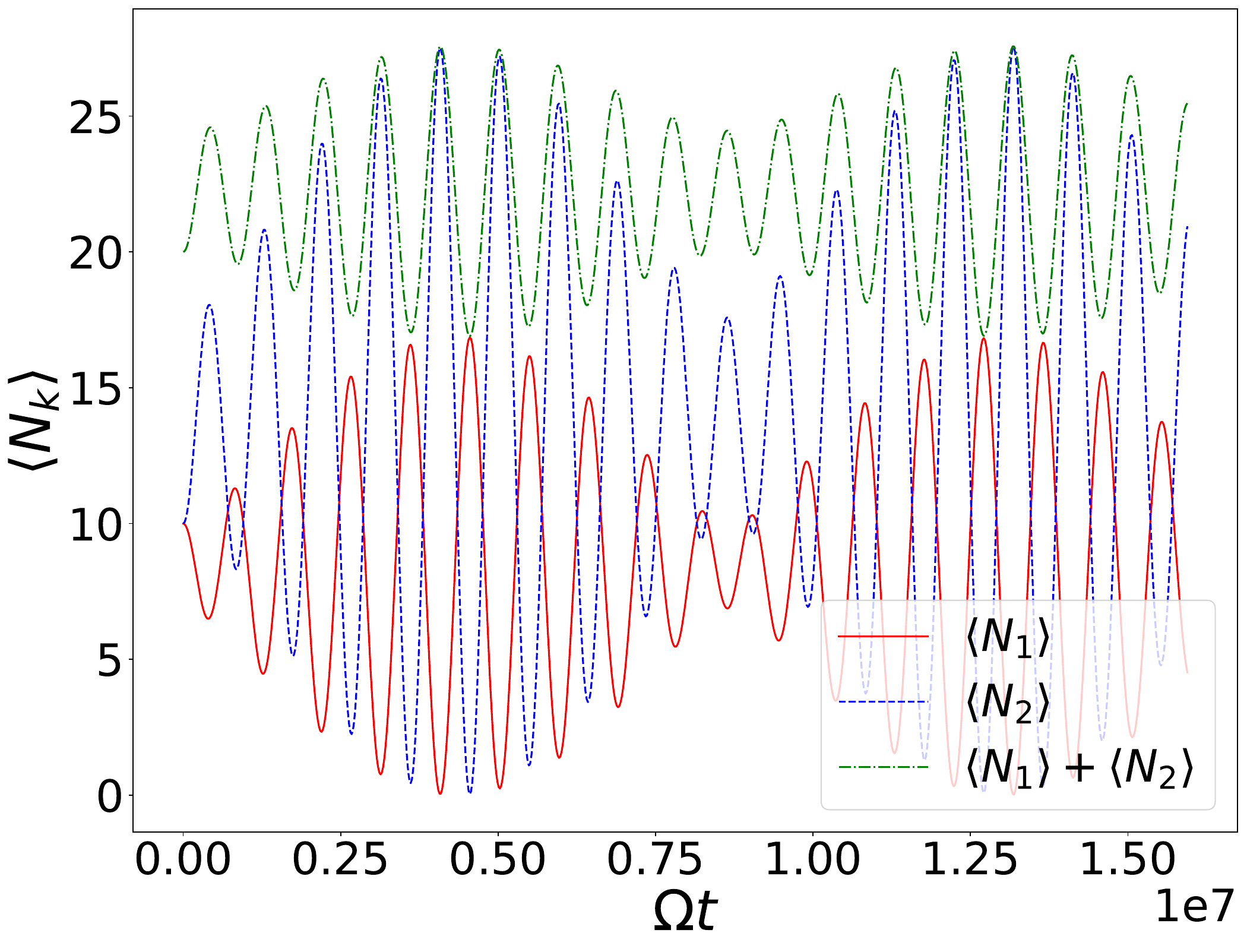}
        \caption{Expected photon number when modes $\omega_1$ and $\omega_2$ are in additive-coupling by driving $\chi$. The data do not correspond to simulations but to expected results from MSA. The cavity was initialised with 10 photons in each coupled mode. $\epsilon=0.01$. }
        \label{fig:batidos}
\end{figure}

Notice that in each mode two sine functions with different frequencies are added up when the respective mode starts with at least one photon. This would produce, for close enough frequencies, beats-like oscillations. In FIG. \ref{fig:batidos} a typical $\chi$-driven additive-coupling is shown to visualize the expected solutions, although no numerical simulations were computed for this scenario. Beats might be harder to identify or non existent if one of the modes has initially no photons, or if there is a sufficiently high mismatch between modes frequencies or initial photon number. We see that oscillations occur, even in the total number of photons. This is the only observed set of solutions where total photon number goes below the initial total photon number, meaning some of the photons were destroyed. It is unclear if this behaviour is to be expected, as MSA is no longer valid at the plotted timescales ($10^3\simeq\epsilon^2\Omega t \cancel{\ll}1$).

\subsection{Difference coupling}
\begin{figure}[H]
         \centering
         \includegraphics[width=.9\columnwidth]{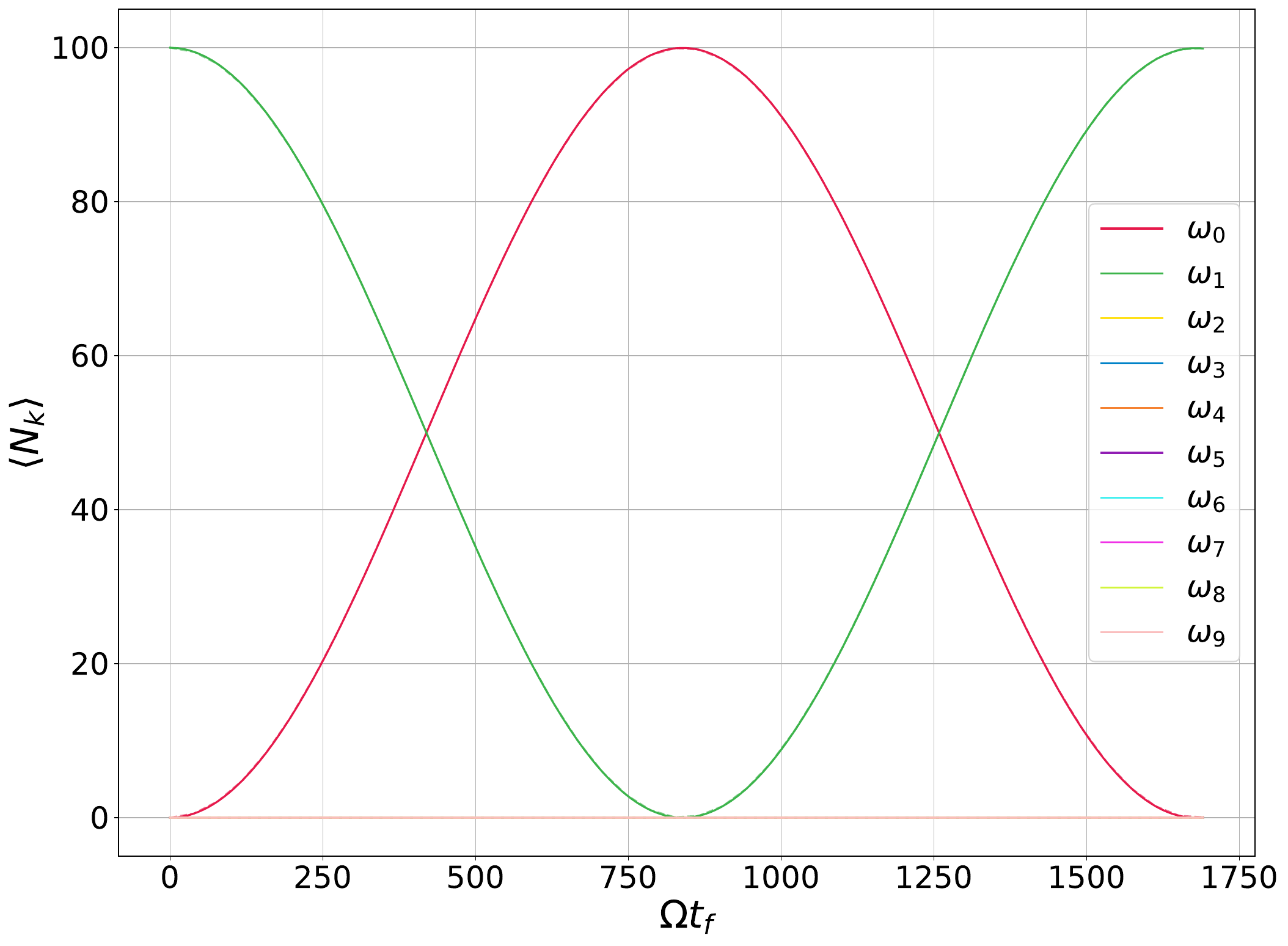}
        \caption{Expected photon number for each cavity mode as a function of time. The dotted lines correspond to \textit{Multiple Scale Analysis} solutions. The system was trimmed at the first $10$ modes, with a cavity with $\Delta L_0/L_0=0.44$, $\chi_0/L_0=0.5$, $v_0L_0=0$ and of unit length. The asymmetry $\Delta L$ was tuned with $\epsilon=0.01$ and $\Omega=\omega_1-\omega_0$.}
        \label{fig:ch4_oscilacion_limpia}
\end{figure}

When driven with a difference coupling, the photon number behaviour changes drastically. It can be shown that photon number not only doesn't increase exponentially but it remains constant. For $\xi_\chi$, we have:
\begin{align}
    \left\langle N_l\right\rangle(t)&=N_n^{(0)}+\left(N_l^{(0)}-N_n^{(0)}\right)\cos^2(\epsilon\gamma\Gamma t),\\
    \left\langle N_n\right\rangle(t)&=N_n^{(0)}+\left(N_l^{(0)}-N_n^{(0)}\right)\sin^2(\epsilon\gamma\Gamma t),\\
    \left\langle N\right\rangle(t)&=N_l^{(0)}+N_n^{(0)},
\end{align}
with \begin{align}
    \gamma&\equiv\frac{\omega_n^2-\omega_l^2}{4\sqrt{\omega_n\omega_l}},\\
    \Gamma&\equiv\sum_{r\neq \chi,\dot{\chi}} r_0\xi_r\beta_{nl}^{(r)},\\
    \Lambda&\equiv\chi_0\xi_\chi\beta_{nl}^{({\chi})}\frac{\omega_l}{\omega_n}.
\end{align}
We see that the total number of photons $N$ is constant over time and depends only on the initial state of the cavity. Also, oscillations are only to be expected if the number of photons in the coupled cavity modes is initially different (i.e. $N_l^{(0)}\neq N_n^{(0)}$). The energy of the cavity does not remain constant, as one can increase/decrease the frequency of the photons by coupling two modes (effectively increasing/decreasing the energy of the cavity). FIG. \ref{fig:ch4_oscilacion_limpia} shows that we can start with 100 photons of $\omega_1$ frequency and convert all of them to photons of $\omega_0$ frequency, reducing the energy of the cavity but keeping the total photon number constant.

\begin{figure}[H]
         \centering
         \includegraphics[width=.9\columnwidth]{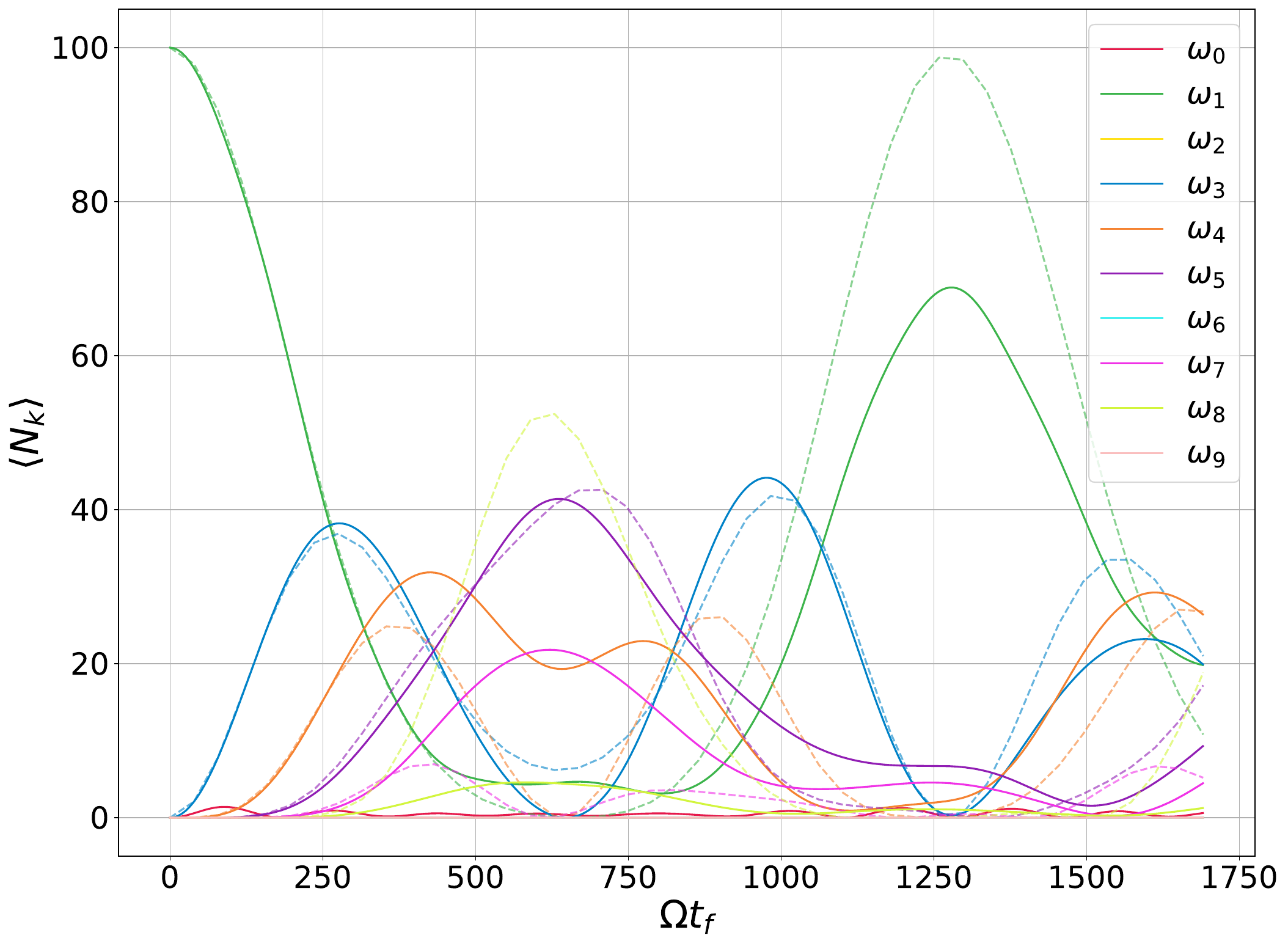}
        \caption{Expected photon number for each cavity mode as a function of time. The dotted lines correspond to \textit{Multiple Scale Analysis} solutions. The system was trimmed at the first $10$ modes, with a cavity with $\Delta L_0/L_0=0.44$, $\chi_0/L_0=5$, $v_0L_0=200$ and of unit length. The asymmetry $\Delta L$ was tuned with $\epsilon=0.01$ and $\Omega=\omega_3-\omega_0$.}
        \label{fig:ch4_oscilacion_muchos_acoplados}
\end{figure}

In FIG. \ref{fig:ch4_oscilacion_limpia} we see that the MSA solutions perfectly match our numerical solutions, even beyond what is expected. But that is not always the case, as the system is sensitive to non-zero detunings. FIG. \ref{fig:ch4_oscilacion_muchos_acoplados} shows many coupled modes, and a small detuning was allowed in the dotted MSA solutions. 

 \subsection{Detuning}

As discussed in the previous sections, the detuning from the driving frequencies plays a key role in the photon generation due to DCE in our cavity. To better understand the effect of the detuning, we shall consider a cavity of unit length, $\Delta L_0/L_0=0.44$, $\chi_0/L_0=0.5$, $v_0L_0=0$, with the asymmetry driven around the $\Omega_0=2\Omega_3$ parametric resonance frequency. The actual driving frequency $\Omega$ introduces a small ($|\Delta\Omega|/\Omega_0\ll1$) detuning such that $\Omega=\Omega_0+\Delta \Omega$. We also choose $\epsilon=0.01$ for all the simulations presented here.

\begin{figure}[H]
     \centering
     \begin{subfigure}[h]{0.99\columnwidth}
         \centering
         \includegraphics[width=\textwidth]{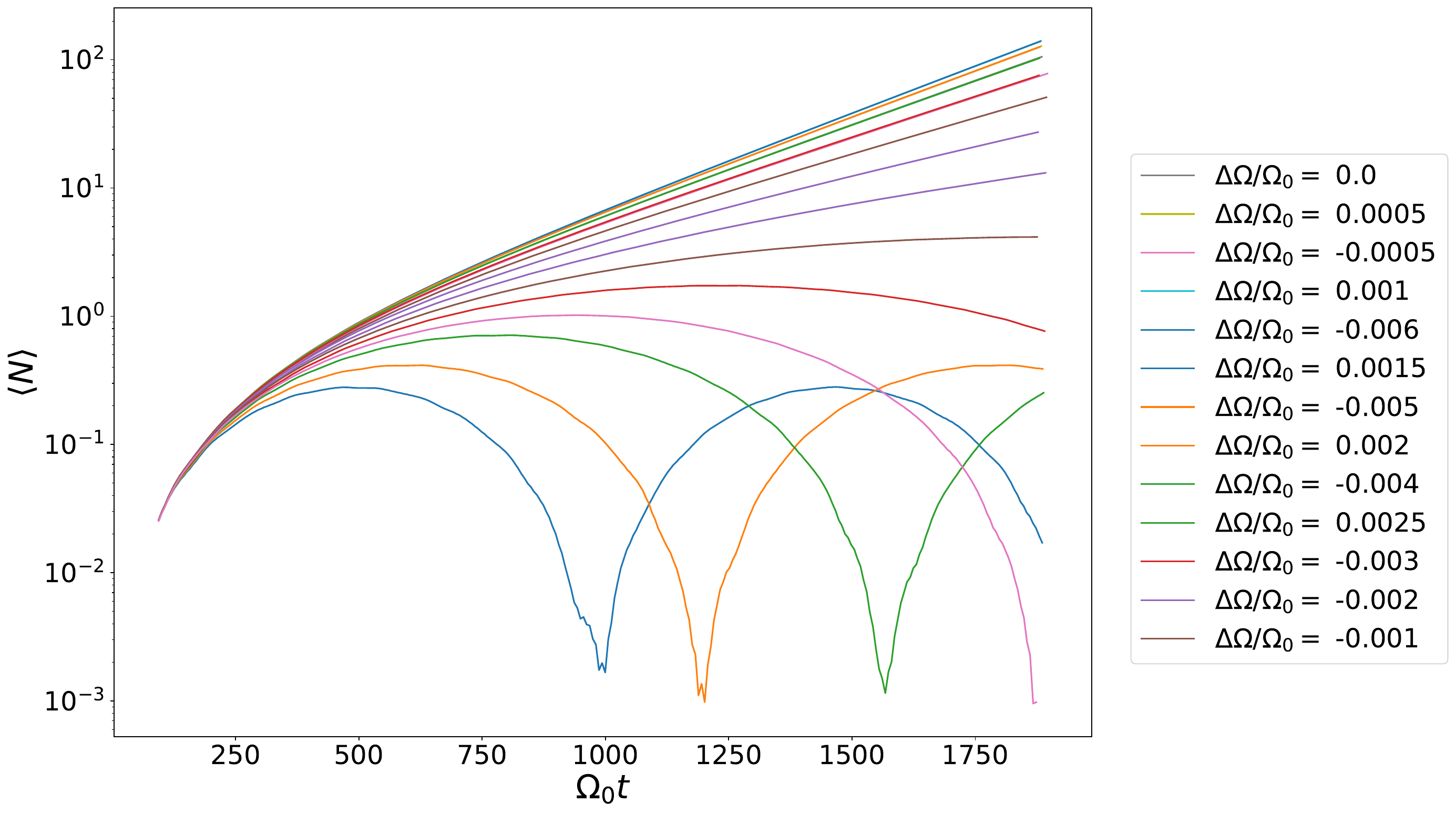}
         \caption{}
         \label{fig:ch4_detuning_b}
     \end{subfigure}
     \\
     \begin{subfigure}[h]{0.99\columnwidth}
         \centering
         \includegraphics[width=\textwidth]{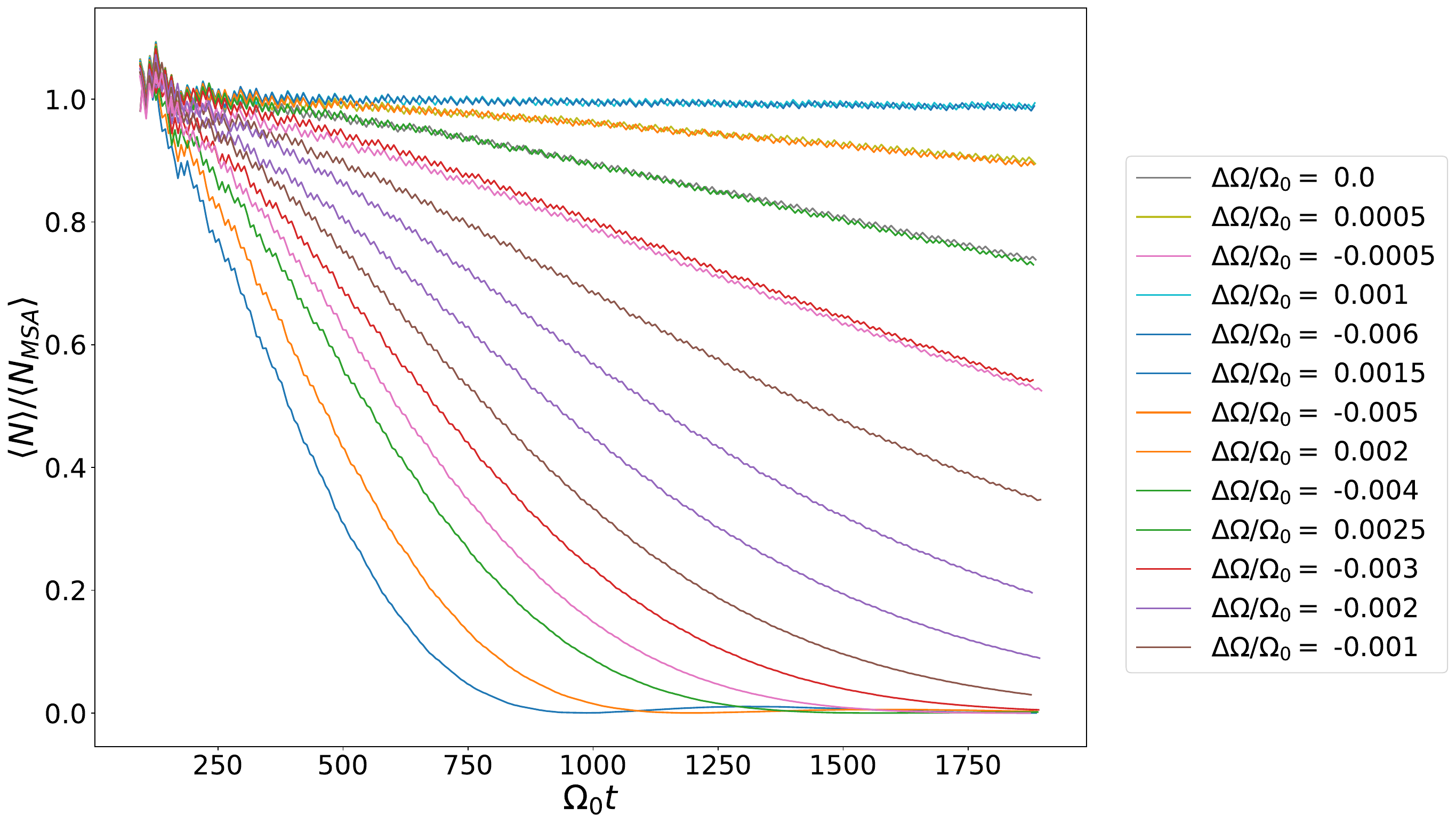}
         \caption{}
         \label{fig:detuning_normalizado}
     \end{subfigure}
        \caption{Expected photon number, for different detunings at the resonant frequency $\Omega_0$. In (b) photon number is normalized against MSA solutions for $\Delta\Omega=0$.}
        \label{fig:ch4_detuning}
\end{figure}

FIG. \ref{fig:ch4_detuning} shows the expected number of photons in time for different values of the detuning $\Delta\Omega$ in the above mentioned scenario. We can clearly see that some detunings (e.g. $\Delta\Omega=0.001$) produce more photons than the zero detuning solution. Still, there is no detuning that produces more photons than those expected by the MSA method (FIG. \ref{fig:detuning_normalizado}). The  $\Delta\Omega=0.001$ curve matches almost perfectly the MSA solution, yet the $\Delta\Omega=0$ curve falls behind it at longer times as already discussed. As it is expected, eventually the detuning becomes too big and no photon generation occurs (we see small oscillations in the numeric solutions).

\begin{figure}[H]
     \centering
     \begin{subfigure}[h]{0.9\columnwidth}
         \centering
         \includegraphics[width=\textwidth]{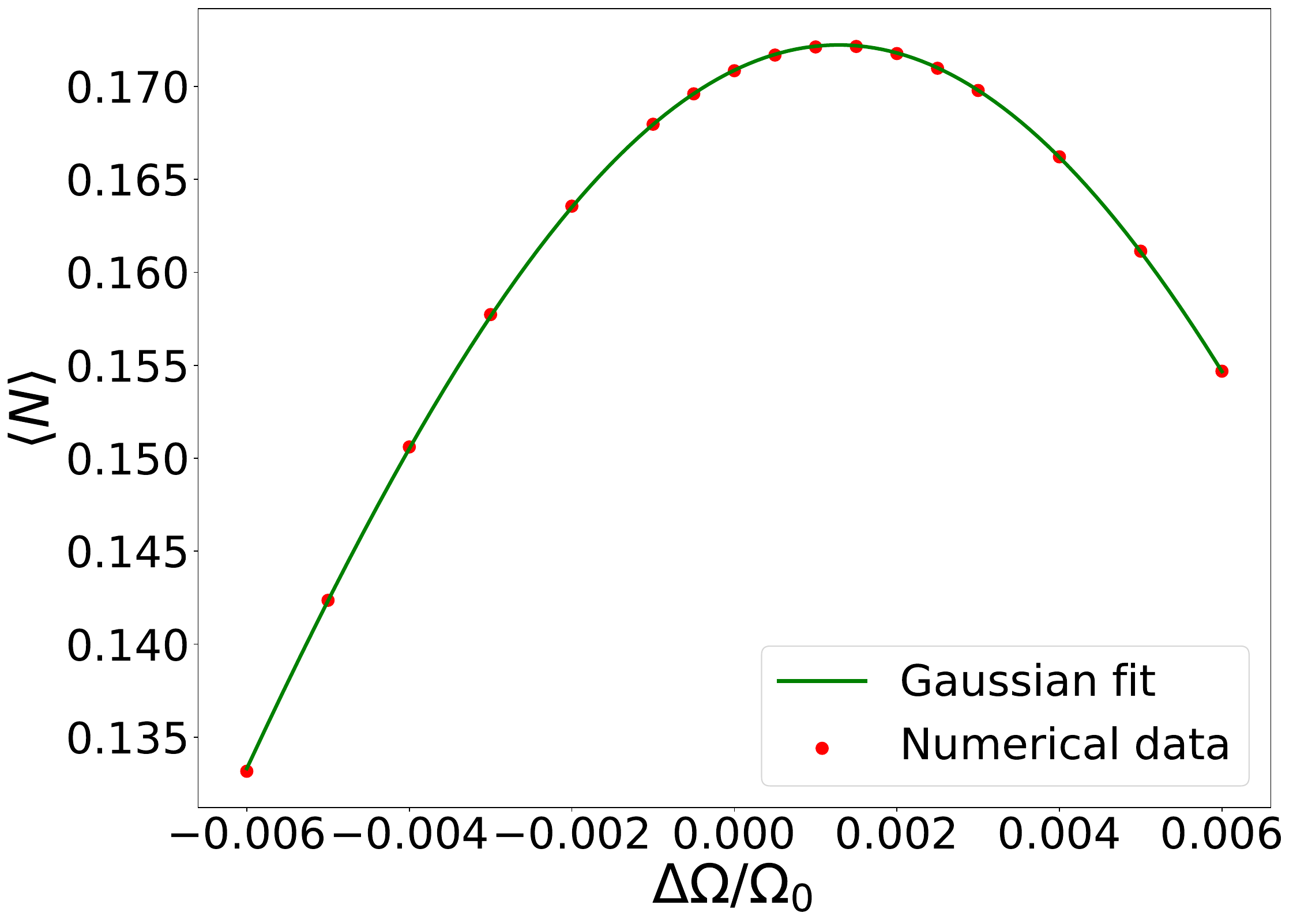}
         \caption{$\Omega_0t\simeq235$}
         \label{fig:ch4_perfil_detuning_a}
     \end{subfigure}
     \\
     \begin{subfigure}[h]{0.9\columnwidth}
         \centering
         \includegraphics[width=\textwidth]{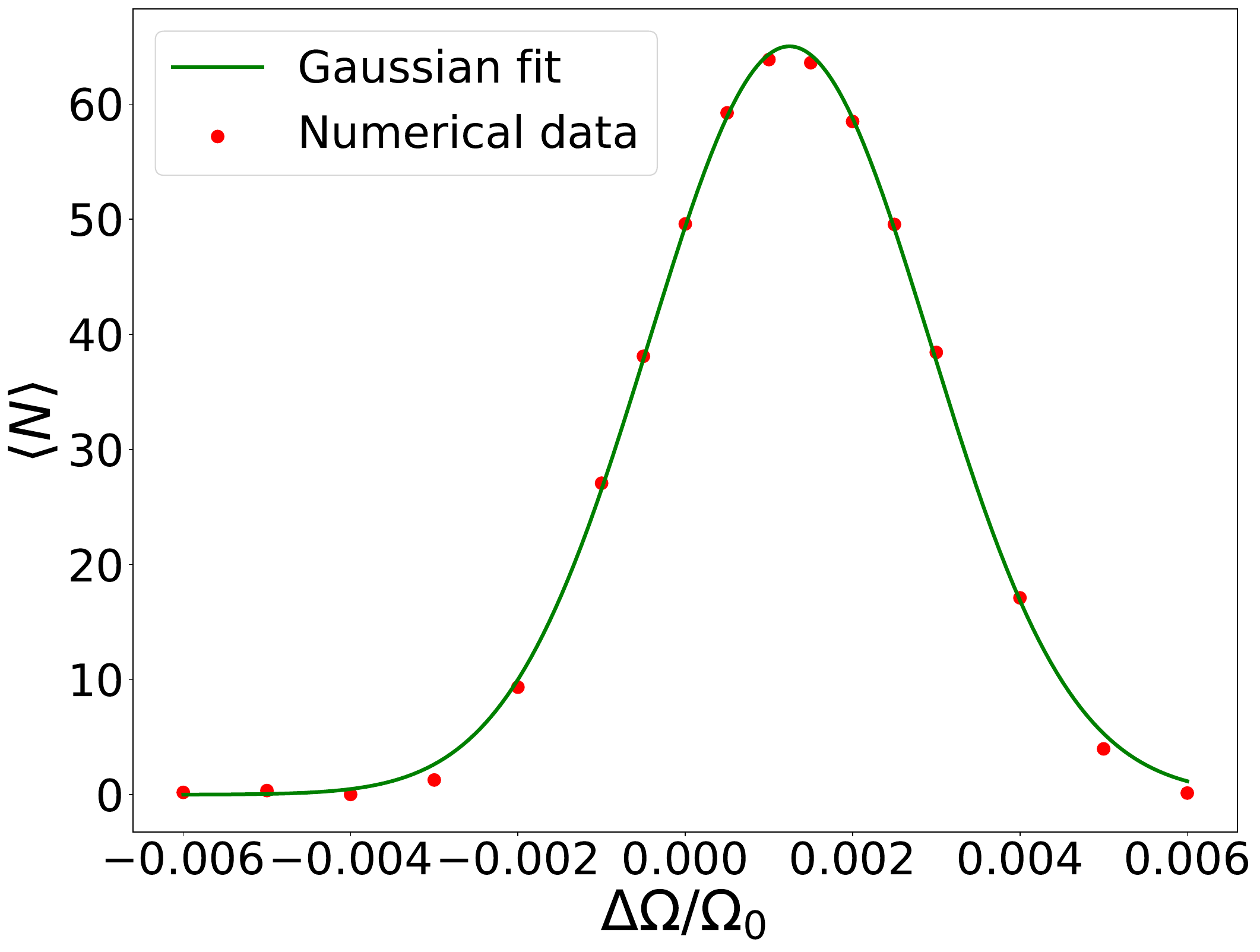}
         \caption{$\Omega_0t\simeq1650$}
         \label{fig:ch4_perfil_detuning_b}
     \end{subfigure}
        \caption{Expectation value of the number operator for photons with frequency $\omega_3$ for different detunings detuning, in a cavity of unit length, $\Delta L_0/L_0=0.44$, $\chi_0/L_0=0.5$, $v_0L_0=0$. The asymmetry $\Delta L$ was tuned with $\epsilon=0.01$ and $\Omega=2\omega_3+\Delta \Omega$. Typical profiles are shown at short times (a),  and long times (b). Gaussian fits are shown in green lines.}
        \label{fig:ch4_perfil_detuning}
\end{figure}

In FIG. \ref{fig:detuning_normalizado} we see that some of the curves stray away from the MSA solution faster and in a non-linear way. When we plot the number of photons in each mode for different time scales (FIG. \ref{fig:ch4_perfil_detuning_a} and FIG. \ref{fig:ch4_perfil_detuning_b}) we see two different profiles. Both of them have their center displaced towards positive values of the detuning, but the width of the distribution is significantly smaller at longer times. 

\begin{figure}[H]
     \centering
     \begin{subfigure}[h]{0.9\columnwidth}
         \centering
         \includegraphics[width=\textwidth]{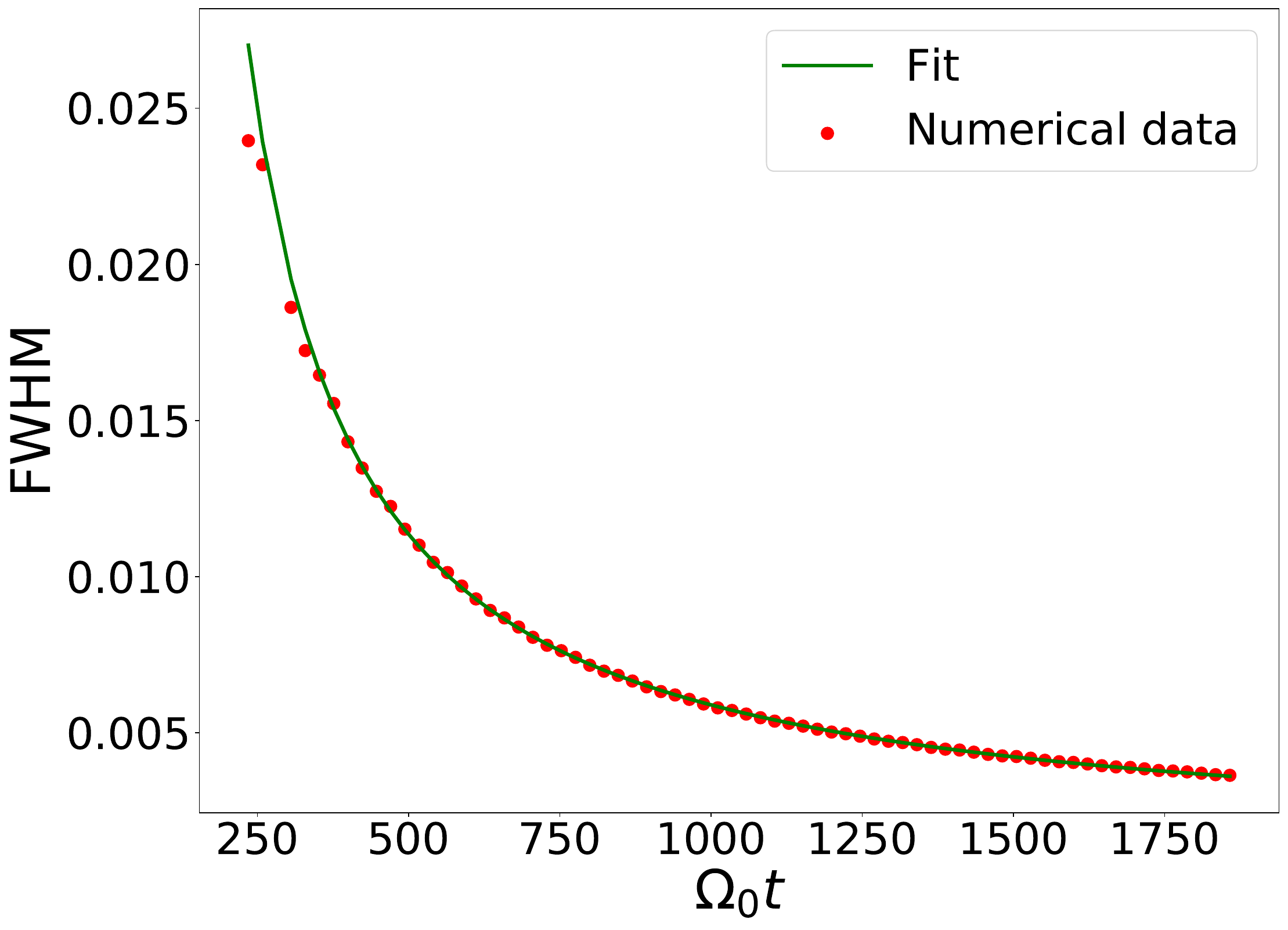}
         \caption{}
         \label{fig:ch4_fwhm}
     \end{subfigure}
     \\
     \begin{subfigure}[h]{0.9\columnwidth}
         \centering
         \includegraphics[width=\textwidth]{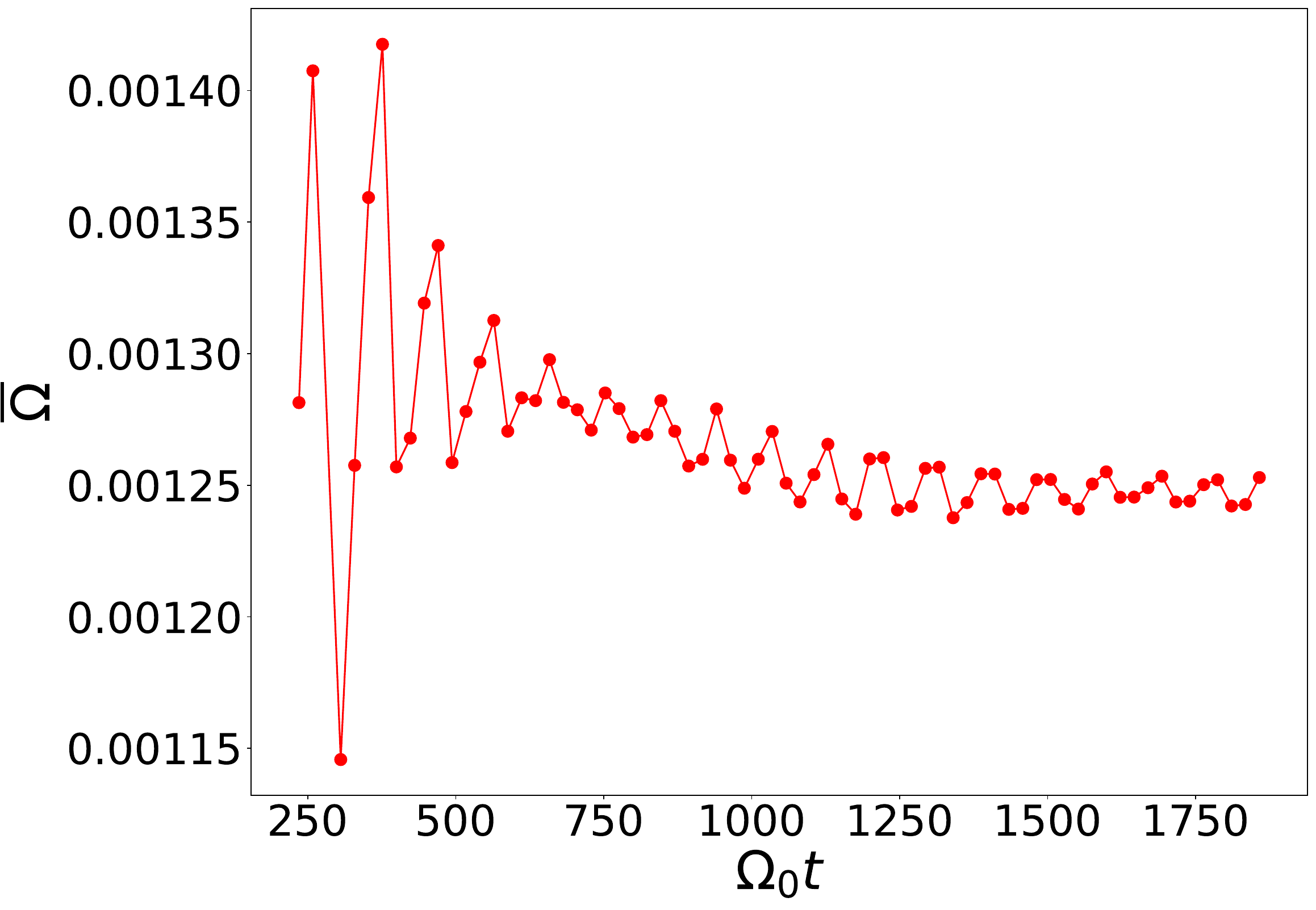}
         \caption{}
         \label{fig:ch4_centros}
     \end{subfigure}
        \caption{(a) FWHM and (b) center, of the Gaussian fits of the detuning profiles. The first points were discarded due to being very noisy. In (a) data was fitted with Eq. \eqref{eq:ch4_ajuste}.}
        \label{fig:ch4_centros_y_fwhm}
\end{figure}

The FWHM $\gamma$ and the center $\overline{\Omega}$ of the Gaussian fit  were computed for different times and are shown in FIG. \ref{fig:ch4_fwhm} and FIG. \ref{fig:ch4_centros}, respectively. In FIG. \ref{fig:ch4_fwhm} we see how the FWHM diminishes over time. This is to be expected due to the exponential nature of the photon generation under the chosen driving conditions, and because the rate was already initially dependent on the detuning. We fitted the width data with a decreasing function: 
\begin{equation}
    \gamma(t)=\frac{A}{t-B}+\gamma_\infty, \label{eq:ch4_ajuste}
\end{equation}
and  obtained an asymptotic width of $\gamma_\infty=0.0015$. 

On the other side, in FIG. \ref{fig:ch4_centros} we see that $\overline{\Omega}$ remains approximately constant throughout all the evolution of the cavity, with some bigger numerical fluctuations at shorter times. We also see that it remains close to $0.001$, as we have already discussed in FIG. \ref{fig:ch4_detuning}.

\begin{figure}[H]
    \centering
    \includegraphics[width=0.99\columnwidth]{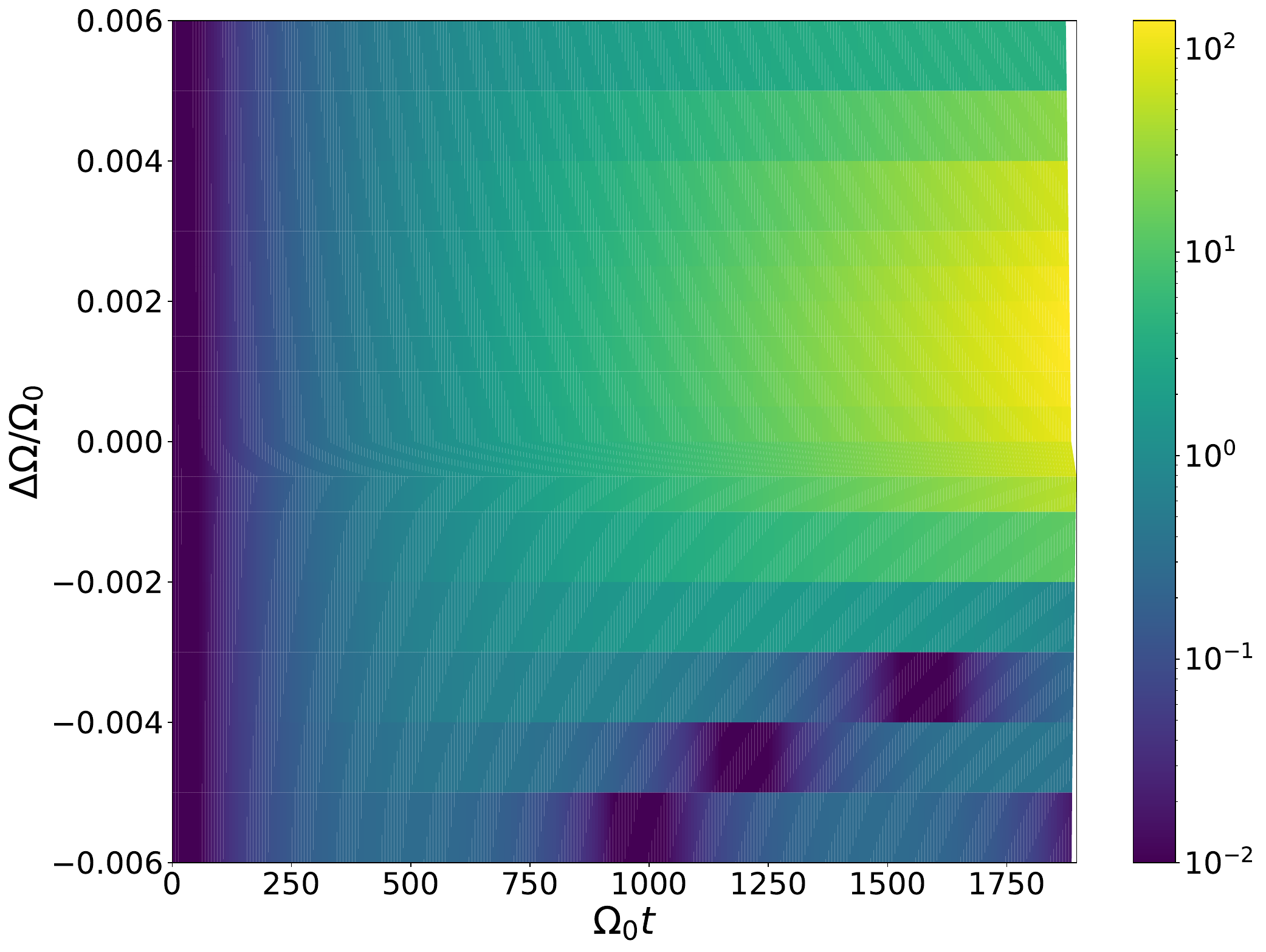}
    \caption{Photon number (in colors) in terms of the dimensionless time and the detuning, in a unit length cavity with $\Delta L_0/L_0=0.44$, $\chi_0/L_0=0.5$, $v_0L_0=0$. The asymmetry $\Delta L$ was tuned with $\epsilon=0.01$ and $\Omega=2\omega_3+\Delta \Omega$.} 
    \label{fig:detuning_heatmap}
\end{figure}

FIG. \ref{fig:detuning_heatmap} presents yet another way of visualizing the previous information. We see that at short times the profile is uniform in the plotted region, which suggests $\gamma>0.01$. As we look at longer times, a region where greater photon production is observed is defined and its contrast increases with time (narrowing the width of the profile as seen in FIG. \ref{fig:ch4_fwhm}). This region is, as discussed in FIG. \ref{fig:ch4_centros}, centered around a positive detuning of $\Delta \Omega/\Omega_0=0.001$. We also see dark regions for $\Delta \Omega/\Omega_0\leq0.003$: these show no true photon generation, but rather numerical oscillations of the solutions. The darkest regions allow us to see the period of the oscillations: the higher the detuning, the smaller the period (until its all dark).

\section{Quantum circuits} \label{sec:circuits}

Performing experiments to observe the DCE in a cavity is virtually impossible, as it requires to keep two mirrors perfectly parallel to each other while moving at GHz frequencies \cite{PhysRevA.100.062516}. In the past years, different proposals have been explored as an alternative to the original mechanical problem \cite{Dodonov96,schaller2002dynamical,kim2006detectability,lambrecht1996motion,sassaroli1994photon}. These consist  on modulating boundary conditions with effective movements that result in parametric amplification of vacuum fluctuations (Parametric DCE, PDCE) \cite{wustmann2013parametric,wilson2011observation,crocce2004model,nation2012colloquium,lozovik1995parametric,uhlmann2004resonant,johansson2013nonclassical, johansson2010dynamical,C.Braggio_2005,SEGEV2007202}. The first successful experiments that resulted in the experimental observation -or quantum simulation- of the DCE were conducted on circuit QED architectures \cite{wilson2011observation,johansson2013nonclassical}.  In this last section, we show a quantum circuit that would allow to effectively tune all four parameters considered in this work.

Waveguides or transmission lines can be used to replace the coupled cavities. Additionally, dc-SQUIDs effectively replace the movement of the mirrors \cite{johansson2009dynamical, wustmann2013parametric,johansson2010dynamical}. We show that an additional dc-SQUID allows for the tuning of the conductivity \cite{velasco2022photon} and that a variable capacitor allows to effectively change the susceptibility of the membrane. 

We define the electric flux
\begin{equation}
    \phi(x,t)=\int_{-\infty}^t V(x,\tilde{t})d\tilde{t},
\end{equation}
where $V$ is the electric potential. We can now write the dc-SQUID Lagrangian density as \cite{garcia2022quantum}:
\begin{equation}
    \mathcal{L}_{\rm dc-SQUID}=\frac{1}{2}(2C_J)(\partial_t\phi)^2+E_J(\varPhi)\cos\left(2\pi\frac{\phi}{\tilde{\varPhi}_0}\right)+I\phi \label{eq:lagrangiano_completo_dcsquid},
\end{equation}
where $\phi$ is the flux difference between its ends,  $\varPhi$ is the external magnetic flux, and $I$ the electric current in the dc-SQUID. $\tilde{\varPhi}_0=(2e)^{-1}$ is the flux quantum.  $E_J$ depends on the external magnetic flux \cite{garcia2022quantum}:
\begin{equation}
    E_J(\varPhi)=2I_C\frac{\tilde{\varPhi}_0}{2\pi}\cos\left(\pi\frac{\varPhi}{\tilde{\varPhi}_0}\right)=E_J(0)\cos\left(\pi\frac{\varPhi}{\tilde{\varPhi}_0}\right),
\end{equation}
where $I_C$ is the critical current of the junction.

Considering $\pi\phi\ll\tilde{\varPhi}_0$ and neglecting $I$, we rewrite Eq. \eqref{eq:lagrangiano_completo_dcsquid} (ignoring a constant term) by expanding the cosine quadratically:
\begin{equation}
    \tilde{\mathcal{L}}_{\rm dc-SQUID}=\frac{1}{2}(2C_J)(\partial_t\phi)^2-\frac{1}{2}\frac{4\pi^2}{{\tilde{\varPhi}_0}^2}E_J(0)\cos\left(\pi\frac{\varPhi}{\tilde{\varPhi}_0}\right)\phi^2.\label{eq:lagrangiano_dcsquid}
\end{equation}
Here we see that the Lagrangian density is quadratic in
$\phi$ and $\partial_t\phi$. Comparing Eq. \eqref{eq:lagrangiano_dcsquid} against Eq. \eqref{eq:densidad_lagrangiana}, we conclude that the capacitance is analogous to electric permittivity, while the non-linear inductance is analogous to electric conductivity. 

\begin{widetext}

\begin{figure}[H]
    \centering
    \includegraphics[width=0.99\textwidth]{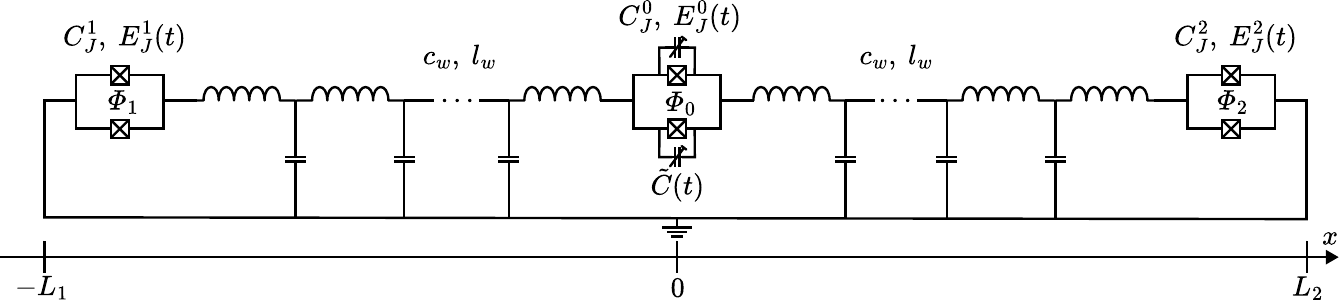}
    \caption{Proposed circuit to perform experimental quantum simulations of the DCE. In each end, a dc-SQUID modulates the effective length of the cavities through external magnetic fluxes $\varPhi_{1,2}$. The central dc-SQUID allows the tuning of $v$ by controlling a third magnetic flux $\varPhi_0$, while the tuning capacitors $\tilde{C}$ let us tune the conductivity. $c_w$ and $l_w$ are the capacitance and inductance per unit length in the coupled transmission lines. Each Josephson junction in each SQUID has an equivalent capacitance $C_J^{0,1,2}$ and a Josephson energy $E_J^{0,1,2}$.} 
    \label{fig:ch5_circuito}
\end{figure}

\end{widetext}

We propose the circuit shown in FIG.  \ref{fig:ch5_circuito} as a cQED equivalent to our cavity with four tunable parameters. It consists of a central dc-SQUID (at $x=0$) coupling two transmission lines (cavities) terminated in another dc-SQUID (mirrors). On each SQUID an external independent magnetic flux is applied, and a variable capacitance is connected in parallel to the central SQUID. 

EM waves propagate inside the transmission lines at speed:
\begin{equation}
    v_w=\frac{1}{\sqrt{l_wc_w}},
\end{equation}
where $l_w$ y $c_w$ are their inductance and capacitance per unit length, and are chosen identical for both cavities. They are moreover time-independent. 

Transmission lines replace cavities, and their Lagrangian density (in the continuum) is:
\begin{equation}
    \mathcal{L}=\frac{1}{2}\left[C(\partial_t\phi)^2-\frac{1}{l_w}(\partial_x\phi)^2\right]-\frac{E}{2\varphi_0}\phi^2, \label{eq:lagrangiano_circuito}
\end{equation}
where
\begin{align}
    \varphi_0&=\frac{\tilde{\varPhi}_0}{2\pi},\\
    C(t)&=c_w+2\left(C_J^{0}+\tilde{C}(t)\right)\delta(x)+2C_J^{1}\delta\left(x+L_1\right)\notag \\ &~~~~+2C_J^{2}\delta\left(x-L_2\right),\\
    E(t)&=E_J^{0}\cos\left(\frac{\varPhi_0}{2\varphi_0}\right)\delta\left(x\right)+E_J^{1}\cos\left(\frac{\varPhi_1}{2\varphi_0}\right)\delta\left(x+L_1\right)\notag\\&~~~~+E_J^{2}\cos\left(\frac{\varPhi_2}{2\varphi_0}\right)\delta\left(x-L_2\right).
\end{align}
In the above expressions, $E_J^{i}=E_J^{i}(0)$. From now on we will use this simplified notation, such that $E_J^{i}$ and $C_J^{i}$ are constants.

For the Lagrangian density in Eq. \eqref{eq:lagrangiano_circuito} the Euler-Lagrange equation reads:
\begin{equation}
    C\partial_t^2\phi+\dot{C}\partial_t\phi-\frac{1}{l_w}\partial_x^2\phi+\frac{E}{\varphi_0}\phi=0. \label{eq:euler_circuito}
\end{equation}
If we restrict ourselves to the transmission lines then $C=c_w$, $\dot{C}=0$, $E=0$ and Eq. \eqref{eq:euler_circuito} is reduced to a propagating wave equation with speed $v_w<1$:
\begin{equation}
    \partial_t^2\phi-v_w^2\partial_x^2\phi=0,
\end{equation}
as expected in analogy to Eq. \eqref{eq: ec_ondas}.

If we instead restrict ourselves to the central dc-SQUID and choose $C_J^0=0$ the Euler-Lagrange equation imposes:
\begin{align}
    &\frac{1}{v_w^2}\left(1+\frac{2\tilde{C}(t)}{c_w}\delta(x)\right)\partial_t^2\phi(t,x)+\frac{1}{v_w^2}\frac{2\dot{\tilde{C}}(t)}{c_w}\delta(x)\partial_t\phi(t,x)\notag \\&-\partial_x^2\phi+\frac{1}{v_w^2}\frac{E_J^0}{c_w\varphi_0}\cos\left(\frac{\varPhi_0}{2\varphi_0}\right)\delta(x)\phi(t,x)=0,
\end{align}
and by integrating this equation we can obtain the discontinuity in $\partial_x\phi$, analogous to Eq.  \eqref{eq:boundary4}. This suggests that in the circuit the total capacitance plays the role of the electric permittivity $\epsilon$ in the cavity, the variable capacitance is equivalent to the membrane's susceptibility $\chi$ and the inductive energy to the electric conductivity $v$:
\begin{align}
     \chi(t)&\longleftrightarrow \frac{1}{v_w^2}\frac{2\tilde{C}(t)}{c_w}, \\ 
     v(t)  &\longleftrightarrow \frac{1}{v_w^2}\frac{E_J^0}{c_w\varphi_0}\cos\left(\frac{\varPhi_0}{2\varphi_0}\right).
\end{align}
This proves the central SQUID presents a way to effectively tune both parameters of the dielectric membrane.

The remaining boundary conditions have been thoroughly discussed in other works \cite{velasco2022photon,wustmann2013parametric,johansson2010dynamical}. It can be shown that the effective lengths of the cavities are:

\begin{align}
    L_1(t)\longleftrightarrow L_1+\frac{1}{v_w^2}\frac{E_J}{c_w\varphi_0}\cos\left(\frac{\varPhi_{1}(t)}{2\varphi_0}\right),\\
    L_2(t)\longleftrightarrow L_2+\frac{1}{v_w^2}\frac{E_J}{c_w\varphi_0}\cos\left(\frac{\varPhi_{2}(t)}{2\varphi_0}\right),
\end{align}
where 
\begin{align}
    \pm\partial_x&\phi(t,\mp L_{1,2})=\notag\\&=\frac{1}{v_w^2}\partial_t^2\phi(t,\mp L_{1,2})+\frac{1}{v_w^2}\frac{E_J}{c_w\varphi_0}\cos\left(\frac{\varPhi_{1,2}}{2\varphi_0}\right)\phi(t,\mp L_{1,2}).
\end{align}
\begin{align}
    \delta d_{1,2}=\frac{1}{v_w^2}\frac{E_J}{c_w\varphi_0}\cos\left(\frac{\varPhi_{1,2}}{2\varphi_0}\right).
\end{align}
For this, we have considered $E_J^1=E_J^2=E_J$ and $C_J^1=C_J^2=0$

If $\varPhi_1(t)=\varPhi_2(t)=\varPhi(t)$ (in-phase fluxes), the asymmetry of the cavity remains constant and equal to its initial value $\Delta L=L_1-L_2$, while the total length can be tuned:
\begin{equation}
    L(t)=L_1+L_2+\frac{2}{v_w^2}\frac{E_J}{c_w\varphi_0}\cos\left(\frac{\varPhi(t)}{2\varphi_0}\right).
\end{equation}
If instead $\varPhi_1(t)=\varPhi_2(t)+2\pi\varphi_0=\varPhi(t)$ (fluxes in counterphase), then the total length $L_1+L_2$ remains constant, while the asymmetry can be tuned as:
\begin{align}
    \Delta L (t)=L_1-L_2+\frac{2}{v_w^2}\frac{E_J}{c_w\varphi_0}\cos\left(\frac{\varPhi(t)}{2\varphi_0}\right).
\end{align}
In arbitrary configurations, both parameters will depend on time and can be controlled independently.

We have shown that the  three magnetic fluxes $\varPhi_{0,1,2}$ and the variable capacitances $\tilde{C}$ allow all four cavity parameters to be tuned. Tuning magnetic fields can be done at much higher frequencies than the current limitations in mechanical systems, and the high sensibility of dc-SQUIDs to external fluxes permits working with low magnetic fields. What presents a challenge is finding the right variable capacitance, as we would want to have something fast but also non dissipative. Varicap diodes might be considered but also recent works \cite{Riwar_capacitores1,Riwar_capacitores2} propose discrete variable capacitances using superconducting material in quantum circuits architectures by exploiting topological properties of Josephson junctions. This could be an interesting alternative to more traditional variable capacitances for our circuit.

\section{Conclusion} \label{sec:conclusiones}

In this work we studied photon generation by means of the Dynamical Casimir Effect (DCE) in a one-dimensional double cavity divided by a dielectric membrane. We consider the tuning of all four parameters of the cavity: the total length ($L$), the difference between the right and left cavities' length (the asymmetry $\Delta L$), the electric susceptibility ($\chi$) and conductivity ($v$) of the dielectric membrane. We considered simultaneous and individual tunings, with independent frequencies, amplitudes and relative phases ($0$ or $\pi$). 

When looking at the energy spectrum of the static cavity, distinct behaviours where identified: high-susceptibility and high-conductivity regimes. We showed that this regimes are divided by a transition zone clearly identifiable by the critical frequency $\omega_c=\sqrt(v/\chi)$, where the highest energy shifts are to be expected. The most interesting difference between these regimes is the existence (prohibition) of a lowest energy mode $\omega_0$ in the high-susceptibility (high-conductivity) regime at low frequencies, $\pi/L>\omega_c$ ($\pi/L<\omega_c$). Avoided crossings were observed in all cavity spectra, regardless of the specific configurations. Asymptotic ($\chi\to\infty, ~~v\to\infty$) solutions were also discussed. 

When considering the instantaneous eigenfrequencies of the dynamic cavity, we showed that they acquire a small non-zero imaginary part if and only if the susceptibility $\chi$ is time-dependent. This effect is neglectable in some cases where other parameters are also being tuned. 

The localization degree $g$ was studied for arbitrary configurations. By choosing appropriate values of $\Delta L$ localized-modes density can be altered (increasing it in the longest cavity and thus decreasing it in the other cavity). The remaining $\chi$ and $v$ must be chosen accordingly to guarantee that the localized modes are indeed solutions of the system. Some modes do not localize at either cavity, but can be forced to localize by changing the cavity parameters and energy spectrum. This is true for all modes but the lowest-energy one $\omega_0<\pi/L$, which was found to be completely de-localized in every cavity configuration (except of course the trivial cases where one of the coupled cavities has zero length).

Numerical solutions for the system where computed, alongside with Multiple Scale Analysis approximations for analytical comparison. Previous known results were recovered for parametric resonance, additive and difference coupling, only when the susceptibility was not being tuned (or its effect was neglectable). The non-zero imaginary part that appears in the instantaneous eigenfrequencies of the dynamic cavity introduces new different behaviours in parametric resonance and additive-coupling. If $\chi$ is resonantly-driven, no exponential growth is observed but rather oscillations in the expected number of photons of the driven mode. This oscillations happen even when the cavity is prepared in a vacuum state, meaning they represent true photon generation. These photons are periodically produced and annihilated, and the process can be understood as an adiabatic shortcut. We note, however, that total photon number does not go below initial total photon number at any point and thus this process alone is not useful for cooling down the cavity. When $\chi$ is tuned by additively coupling two different modes, beats-like oscillations can be observed if parameters are properly chosen. In this scenario, photon number does go below initial photon number.

\section*{Acknowledgments}
We thank F. D. Mazzitelli and C. T. Schmiegelow for valuable discussions and helpful feedback.
This research was funded by Agencia Nacional de
Promoción Científica y Tecnológica (ANPCyT), Consejo
Nacional de Investigaciones Cientıficas y Técnicas
(CONICET), and Universidad de Buenos Aires
(UBA). This research was also funded in part by the Austrian Science Fund (FWF) under grant 10.55776/COE1. For open access purposes, the author has applied a CC BY public copyright license to any author-accepted manuscript version arising from this submission.

\appendix

\section{Multiple Scale Analysis}\label{ap:MSA}

We consider small sinusoidal perturbations in all cavity parameters:
\begin{align}
L(t)&=L_0(1+\epsilon\xi_L\sin(\Omega_Lt))~\label{eq:def_paramL}, \\
\Delta L(t)&=\Delta L_0(1+\epsilon\xi_{\Delta L}\sin(\Omega_{\Delta L}t)),\\
\chi(t)&=\chi_0(1+\epsilon\xi_\chi \sin(\Omega_\chi t)),\\
v(t)&=v_0(1+\epsilon\xi_v\sin(\Omega_vt))\label{eq:def_paramv}
\end{align}
where $\Omega_r$ is the  driving frequency of the $r$ parameter, $\epsilon$ is small ($|\epsilon|\ll1$) and fixes the order of all perturbations.  $\xi_r$ satisfies $|\xi_r|\simeq1$ or $\xi_r=0$ and allows to have different amplitudes and relative phases ($0$ or $\pi$) for each driven parameter. To turn off the driving of a specific parameter, we choose $\xi_r=0$ for that parameter. $\dot{\chi}$ is the last relevant parameter and is easily obtained by differentiating the above expression for $\chi$.

As already discussed, the tuning of the parameters must stop after a finite time $t_f$. We can note the discontinuity of the parameters' first derivatives at $t=0$ as defined in Eq. \eqref{eq:def_paramL} to \eqref{eq:def_paramv}. However, this issue can be easily addressed by adding time dependent functions that rapidly decrease with time, such as:
\begin{equation}
    f_r(t)=-\epsilon r_0 \xi_r \Omega_r t e^{-\alpha t},
\end{equation}
with $\alpha\in\mathbb{R}^{+}$. We may redefine the parameters as:
\begin{equation}
    \tilde{r}(t)=r(t)+f_r(t).
\end{equation}
Because $\alpha$ can be arbitrarily big, the modification disappears arbitrarily fast and is virtually indistinguishable from the original sinusoidal trajectories (except, of course, at $t=0$ and $t=t_f$). Under all these conditions, the MSA solutions will be valid at times $t$ that satisfy simultaneously:
\begin{align}
t\epsilon^2\max_{r=L,\Delta L, \chi, v}\{\xi_r^2\Omega_r\}&\ll1\\
\alpha t&\gg1
\end{align}

\begin{widetext}

We now define a second, slower time scale $\tau \equiv\epsilon t$, and perform first order in $\epsilon$ expansions of the $Q_n^{(m)}$ coefficients. For this,  we shall initially consider $t$ and $\tau$ as independent variables \cite{bender1999advanced}:
\begin{equation}
    Q_n^{(m)}(t)=Q_n^{(m)(0)}(t,\tau)+\epsilon Q_n^{(m)(1)}(t,\tau)+\mathcal{O}(\epsilon^2). \label{eq:taylor_Q}
\end{equation}
From \eqref{eq:taylor_Q} and by the definition of $\tau =\epsilon t$ we may compute the coefficients' derivatives:
\begin{align}
\dot{Q}_n^{(m)}&=\partial_tQ_n^{(m)(0)}(t,\tau)+\epsilon\left(\partial_\tau Q_n^{(m)(0)}(t,\tau)+\partial_t Q_n^{(m)(1)} (t,\tau)\right)+\mathcal{O}(\epsilon^2),\label{eq:qpunto}\\
\ddot{Q}_n^{(m)}&=\partial_t^2Q_n^{(m)(0)}(t,\tau)+\epsilon\left(2\partial_{t\tau}^2 Q_n^{(m)(0)}(t,\tau)+\partial_t^2 Q_n^{(m)(1)} (t,\tau)\right)+\mathcal{O}(\epsilon^2).\label{qpuntopunto}
\end{align}

We can also expand the square of the instantaneous eigenfrequencies:
\begin{equation}
    \omega_n^2(t)=\omega_{n,0}^2+\epsilon \omega_{n,1}^{2}(t)+\mathcal{O}(\epsilon^2),
\end{equation}
where $\omega_{n,0}$ is the $n$-th mode (static) eigenfrequency, and $\omega_{n,1}^{2}(t)$ is given by a Taylor expansion:
\begin{align}
\epsilon\omega_{n,1}^{2}(t)&=\sum_{r=L,\Delta L, \chi, v, \dot{\chi}} \left.(\partial_r\omega_n^2)\right|_0 \Delta{r}\notag\\
\omega_{n,1}^{2}(t)&=2\omega_{n,0}\left[\xi_\chi\chi_0 \Omega_\chi\cos(\Omega_\chi t) \eta_n^{(\dot{\chi})}+\sum_{r=L,\Delta L, \chi, v} \xi_rr_0\sin(\Omega_rt) \eta_n^{(r)} \right].\label{eq:frecuencia_orden1}
\end{align}

At zeroth order in $\epsilon$, Eq. \eqref{eq:ecuacion_Q} is that of a harmonic oscillator:
\begin{align}
\partial_t^2Q_n^{(m)(0)}(t,\tau)+\omega_{n,0}^2Q_n^{(m)(0)}(t,\tau)=0,
\end{align}
yielding a solution:
\begin{equation}
    Q_n^{(m)(0)}=A_n^m(\tau)e^{i\omega_{n,0}t}+B_n^m(\tau)e^{-i\omega_{n,0}t}.\label{eq:sol_ordencero}
\end{equation}

On the other hand, discarding terms of order $\mathcal{O}(\epsilon)^2$:
\begin{equation}
\resizebox{.9\textwidth}{!}{$\displaystyle{
\epsilon\left(2\partial_{t\tau}^2 Q_l^{(m)(0)}+\partial_t^2 Q_l^{(m)(1)} + \omega_{l,0}^{2~(0)} Q_l^{(m)(1)}\right)+\omega_{l,1}^{2}Q_l^{(m)(0)}=-\sum_{n,r} Q_n^{(m)(0)}\ddot{r}\beta_{nl}^{(r)}  + 
2\partial_tQ_n^{(m)(0)}\dot{r}\beta_{nl}^{(r)}}$},\label{eq:ordenuno}
\end{equation}
Eq. \eqref{eq:ordenuno} is evidently a set of infinite coupled equations, and thus the system will need to be trimmed in order to be solved in the more general situations. This is achieved by considering only the first $N$ modes in the cavity in every sum. 

To avoid secular terms, we impose that the terms multiplied by complex exponential functions $\exp{\pm i\omega_l}$ cancel out (or are equal in the LHS and RHS of Eq. \eqref{eq:ordenuno}). Ignoring $Q_l^{(m)(1)}$:
\begin{align}
-2\partial_{t\tau}^2 Q_l^{(m)(0)}&=\notag\\
&2\omega_l\left[\xi_\chi\chi_0\Omega_\chi \cos(\Omega_\chi t)\eta_l^{(\dot{\chi})}+\sum_{r=L,\Delta L, \chi, v} \xi_rr_0\sin(\Omega_rt)\eta_l^{(r)} \right]Q_l^{(m)(0)}+\notag\\ 
&\sum_{n} -\chi_0\xi_\chi\Omega_\chi^3 \beta_{nl}^{(\dot{\chi})} \cos(\Omega_\chi t) Q_n^{(m)(0)} - 
2\chi_0\xi_\chi\Omega_\chi^2\beta_{nl}^{(\dot{\chi})}\sin(\Omega_\chi t)\partial_tQ_n^{(m)(0)} \notag\\ 
&\sum_{n,r\neq \dot{\chi}} -r_0\xi_r\Omega_r^2\beta_{nl}^{(r)}\sin(\Omega_rt)Q_n^{(m)(0)}  + 
2r_0\xi_r\Omega_r\beta_{nl}^{(r)}\cos(\Omega_rt)\partial_tQ_n^{(m)(0)}.\label{eq:secularesnt}
\end{align}

From Eq. \eqref{eq:sol_ordencero} we get the derivatives of $Q_n^{(m)(0)}$:
\begin{align}
    \partial_{t}Q_n^{(m)(0)}&=i\omega_nA_n^m(\tau)e^{i\omega_nt}-i\omega_nB_n^m(\tau)e^{-i\omega_nt} \notag\\ 
-2\partial_{t\tau}^2Q_l^{(m)(0)}&=-2i\omega_l\left(\frac{dA_l^m}{d\tau}e^{i\omega_lt}-\frac{dB_l^m}{d\tau}e^{-i\omega_lt}\right),
\end{align}
and by introducing them and Eq. \eqref{eq:sol_ordencero} into Eq. \eqref{eq:secularesnt} we obtained another coupled system for the slow varying coefficients $A_n^m(\tau)$ and $B_n^m(\tau)$:

\begin{align}
&-2i\omega_l\frac{dA_l^m}{d\tau}=\notag\\
&+\omega_l\xi_\chi\chi_0\Omega_\chi 
B_l^m \eta_l^{(\dot{\chi})} \delta(\Omega_\chi-2\omega_l) 
-i\omega_l\sum_{r\neq \dot{\chi}} \xi_rr_0
B_l^m \eta_l^{(r)}\delta(\Omega_r-2\omega_l)\notag\\ 
&-\sum_{n} \chi_0\xi_\chi\Omega_\chi^3 \beta_{nl}^{(\dot{\chi})} \frac{1}{2}\left\{A_n^m \left[\delta(\Omega_\chi+\omega_n- \omega_l) +\delta(-\Omega_\chi+\omega_n- \omega_l)\right]+
B_n^m \delta(\Omega_\chi-\omega_n- \omega_l)\right\}\notag\\
&-\sum_{n} 2\chi_0\xi_\chi\Omega_\chi^2\beta_{nl}^{(\dot{\chi})}\frac{i\omega_n}{2i}\left\{A_n^m \left[\delta(\Omega_\chi+\omega_n- \omega_l)-\delta(-\Omega_\chi+\omega_n- \omega_l)\right]-
B_n^m \delta(\Omega_\chi-\omega_n- \omega_l)\right\} \notag\\ 
&-\sum_{n,r\neq \dot{\chi}} r_0\xi_r\Omega_r^2\beta_{nl}^{(r)}\frac{1}{2i}\left\{A_n^m \left[\delta(\Omega_r+\omega_n- \omega_l)-\delta(-\Omega_r+\omega_n- \omega_l)\right]+
B_n^m \delta(\Omega_r-\omega_n- \omega_l)\right\} \notag \\
&+\sum_{n,r\neq \dot{\chi}}2r_0\xi_r\Omega_r\beta_{nl}^{(r)}\frac{i\omega_n}{2}\left\{A_n^m \left[\delta(\Omega_r+\omega_n- \omega_l) +\delta(-\Omega_r+\omega_n- \omega_l)\right]-
B_n^m \delta(\Omega_r-\omega_n- \omega_l)\right\}\label{eq:MSA_A}
\end{align}

\begin{align}
&2i\omega_l\frac{dB_l^m}{d\tau}=\notag \\
&+\omega_l\xi_\chi\chi_0 \Omega_\chi A_l^m \eta_l^{(\dot{\chi})}\delta(\Omega_\chi-2\omega_l)
+i\omega_l\sum_{r\neq \dot{\chi}} \xi_rr_0A_l^m \eta_l^{(r)} \delta(\Omega_r-2\omega_l)\notag\\ 
&-\sum_{n} \chi_0\xi_\chi\Omega_\chi^3 \beta_{nl}^{(\dot{\chi})} \frac{1}{2}\left\{A_n^m \delta(\Omega_\chi-\omega_n- \omega_l)+
B_n^m \left[\delta(\Omega_\chi-\omega_n+ \omega_l) +\delta(-\Omega_\chi-\omega_n+ \omega_l) \right]\right\}\notag\\
&+\sum_{n} 2\chi_0\xi_\chi\Omega_\chi^2\beta_{nl}^{(\dot{\chi})}\frac{i\omega_n}{2i}\left\{A_n^m \delta(\Omega_\chi-\omega_n- \omega_l)+
B_n^m \left[\delta(\Omega_\chi-\omega_n+ \omega_l)-\delta(-\Omega_\chi-\omega_n+ \omega_l)\right]\right\} \notag\\
&-\sum_{n,r\neq \dot{\chi}} r_0\xi_r\Omega_r^2\beta_{nl}^{(r)}\frac{1}{2i}\left\{-A_n^m \delta(\Omega_r-\omega_n- \omega_l)+
B_n^m \left[\delta(\Omega_r-\omega_n+ \omega_l)-\delta(-\Omega_r-\omega_n+ \omega_l)\right]\right\}  \notag\\
&+\sum_{n,r\neq \dot{\chi}}2r_0\xi_r\Omega_r\beta_{nl}^{(r)}\frac{i\omega_n}{2}\left\{A_n^m \delta(\Omega_\chi-\omega_n- \omega_l)-
B_n^m \left[\delta(\Omega_r-\omega_n+ \omega_l) +\delta(-\Omega_r-\omega_n+ \omega_l) \right]\right\}\label{eq:MSA_B}
\end{align}
In the above expressions, $\delta(x)$ is the Kronecker delta function.

The problem is completed by the initial conditions for $Q$ and $\dot{Q}$ imposed by Eq. \eqref{eq:cond_inicial_Q} and \eqref{eq:cond_inicial_Qdot}. The Bogoliubov transformations (Eq. \eqref{eq:bogo1} and \eqref{eq:bogo2}) allow us to obtain the equivalent border conditions for $A$ and $B$ \cite{velasco2022photon}:
\begin{align}
    A_n^{(m)}(0)&=0\label{eq:ch3_cond_inicial_A}\\
    B_n^{(m)}(0)&=\frac{\delta_{mn}}{\sqrt{2\omega_n}}\label{eq:ch3_cond_inicial_B}
\end{align}

\section{Properties of the $\beta$ coefficients}
 $\beta_{nl}^{(r)}$ is defined by Eq. \eqref{eq:beta}. First, we assume $r$ is  $L$, $\Delta L$ or $v$ (and not $\chi$ or its derivative). Then by using the definition of the inner product (Eq. \eqref{eq:producto_interno}) we may write:
\begin{align}
    \beta_{nl}^{(r)}=(\partial_r \varPhi_n, \varPhi_l)&=\int^{L_2}_{-L_1}dx ~(\partial_r\varPhi_n(x))\varPhi_l(x)\left(1+\delta(x)\chi\right) \notag \\
    &=\int^{L_2}_{-L_1}dx ~\left[\partial_r(\varPhi_n(x)\varPhi_l(x))-\varPhi_n(x)(\partial_r\varPhi_l(x))\right]\left(1+\delta(x)\chi\right) \notag \\
    &=-\beta_{ln}^{(r)}+\int^{L_2}_{-L_1}dx ~\partial_r(\varPhi_n(x)\varPhi_l(x))\left(1+\delta(x)\chi\right) \notag \\
    &=-\beta_{ln}^{(r)}+\int^{L_2}_{-L_1}dx ~\partial_r(\varPhi_n(x)\varPhi_l(x)\left(1+\delta(x)\chi\right))    ~. \label{eq:apC_1}
\end{align}
We then use the Leibniz rule for differentiation under the integral symbol:
\begin{align}
\int^{L_2}_{-L_1}dx ~\partial_r(\varPhi_n(x)\varPhi_l(x)\left(1+\delta(x)\chi\right))&=\partial_r \int^{L_2}_{-L_1}dx ~\varPhi_n(x)\varPhi_l(x)\left(1+\delta(x)\chi\right)\notag \\
&\phantom{==}-\varPhi_n(-L_1)\varPhi_l(-L_1)\left(1+\delta(-L_1)\chi\right)\partial_r (-L_1)\notag \\
&\phantom{==} +\varPhi_n(L_2)\varPhi_l(L_2))\left(1+\delta(L_2)\chi\right) \partial_r L_2~. \label{eq:apC_leibniz}
\end{align}
The first line in \eqref{eq:apC_leibniz} is $\partial_r (\varPhi_n, \varPhi_l)=0$ due to the set of $\varPhi$ functions being orthonormal. The remaining two lines are also zero due to the boundary conditions imposed by two perfectly conducting mirrors. Then the RHS in \eqref{eq:apC_leibniz} is zero and then Eq. \eqref{eq:apC_1} can be simply written as:
\begin{align}
    \beta_{nl}^{(r)}=-\beta_{ln}^{(r)}~, \label{eq:apC_betanochi}
\end{align}
for $r=L$, $\Delta L$, $v$. 

For $r=\chi$ we must be careful with the permittivity term that is present in the inner product's definition:
\begin{align}
    \beta_{nl}^{(\chi)}&=\int^{L_2}_{-L_1}dx ~(\partial_\chi\varPhi_n(x))\varPhi_l(x)\left(1+\delta(x)\chi\right) \notag \\
    &=\int^{L_2}_{-L_1}dx ~\left[\partial_\chi(\varPhi_n(x)\varPhi_l(x)\left(1+\delta(x)\chi\right))-\varPhi_n(x)(\partial_\chi\varPhi_l(x))\left(1+\delta(x)\chi\right)-\delta(x)\varPhi_n(x)\varPhi_l(x)\right] \notag \\
    &=\int^{L_2}_{-L_1}dx ~\partial_\chi\left[\varPhi_n(x)\varPhi_l(x)\left(1+\delta(x)\chi\right)\right] -\beta_{ln}^{(\chi)}-\varPhi_n(0)\varPhi_l(0)\notag \\
    &=-\beta_{ln}^{(\chi)}-\varPhi_n(0)\varPhi_l(0)  ~.
\end{align}
\end{widetext}
Defining
\begin{equation}
    \sigma_{nl}\equiv \varPhi_n(0)\varPhi_l(0)
\end{equation}
we rewrite the effect of the indexes permutation for the coefficient $\beta_{nl}^{(\chi)}$:
\begin{equation}
    \beta_{nl}^{(\chi)}=-\beta_{ln}^{(\chi)}-\sigma_{nl}~.\label{eq:apC_betachi}
\end{equation}
Of course,  $\sigma_{nl}$ is symmetrical under indexes exchange and thus $\sigma_{nl}=\sigma_{ln}$.

To write the effect of exchanging indexes for the coefficient associated to $\dot{\chi}$, we must first note that we can compute the partial derivatives of the eigenfrequencies with respect to the cavity parameters through the Implicit Function Theorem:
\begin{equation}
     \frac{\partial k_n}{\partial r}=-\frac{\partial F}{\partial r}\left(\frac{\partial F}{\partial k_n}\right)^{-1}~, \label{eq:TFI}
\end{equation}
where $F=0$ is any of all possible functions whose roots are the cavity's eigenfrequencies. It can be shown that the ratios of these derivatives are rather simple for the dielectric membrane parameters:
\begin{align}
    \frac{\left(\partial_\chi k_n\right)}{\left(\partial_{\dot{\chi}}k_n\right)}&=-ik_n~,\label{eq:apB_cociente1} \\ 
    \frac{\left(\partial_\chi k_n\right)}{\left(\partial_{v}k_n\right)}&=-k_n^2~,\label{eq:apB_cociente2} \\ 
    \frac{\left(\partial_{\dot{\chi}} k_n\right)}{\left(\partial_v k_n\right)}&=-ik_n~.\label{eq:apB_cociente3} 
\end{align}

Now, using Eq. \eqref{eq:apB_cociente1} and Eq.   \eqref{eq:eta_definicion}: 
\begin{align}
    \beta_{nl}^{({\chi})}&=(\partial_{{\chi}} \varPhi_n, \varPhi_l)\notag\\&=((\partial_{{\chi}}k_n)\partial_{k_n}\varPhi_n-(\partial_\chi N_n^2)\varPhi_n/(2N_n^2),\varPhi_l)\notag\\
    &=(\eta_n^{(\chi)}\partial_{k_n}\varPhi_n+\nu_n\varPhi_n,\varPhi_l)\notag\\&=\eta_n^{(\chi)}(\partial_{k_n}\varPhi_n,\varPhi_l)+\nu_n(\varPhi_n,\varPhi_l)\notag\\&=\eta_n^{(\chi)}(\partial_{k_n}\varPhi_n,\varPhi_l)+\nu_n\delta_{nl}~, \label{eq:apC_betachi_desarrollo}
\end{align}
where $N_n$ refers to the normalization constant of the $n$-th mode, and
\begin{equation}
    \nu_n\equiv -\frac{\partial_\chi N_n^2}{2N_n^2}.
\end{equation}
In an analogous way we can work on the coefficients $\beta_{nl}^{(v)}$ and $\beta_{nl}^{(\dot{\chi})}$ to obtain, for $n\neq l$:
\begin{align}
    \frac{\beta_{nl}^{(\chi)}}{\beta_{nl}^{(\dot{\chi})}}&=-ik_n\label{eq:apC_relacionbetachichipunto},  \\
    \frac{\beta_{nl}^{(\chi)}}{\beta_{nl}^{(v)}}&=-k_n^2\label{eq:apC_relacionbetachiv}, \\ 
    \frac{\beta_{nl}^{(\dot{\chi})}}{\beta_{nl}^{(v)}}&=-ik_n,
\end{align}
where we now used Eq. \eqref{eq:apB_cociente2} and Eq. \eqref{eq:apB_cociente3}

By putting together Eq. \eqref{eq:apC_betanochi}, Eq. \eqref{eq:apC_betachi} and Eq.  \eqref{eq:apC_relacionbetachiv} we find a simple analytic expression for $\beta_{nl}^{(v)}$:
\begin{align}
-k_n^2\beta_{nl}^{(v)}&=k_l^2\beta_{ln}^{(v)}-\sigma_{nl}, \notag \\
-k_n^2\beta_{nl}^{(v)}&=-k_l^2\beta_{nl}^{(v)}-\sigma_{nl},\notag \\
    \beta_{nl}^{(v)}&=\frac{\sigma_{nl}}{k_n^2-k_l^2}, \label{eq:beta_v_completa}
\end{align}
only valid for  $n\neq l$.

By replacing Eq. \eqref{eq:beta_v_completa} in Eq. \eqref{eq:apC_relacionbetachiv} we can write the non-diagonal terms of $\beta_{nl}^{\chi}$ as:
\begin{equation}
    \beta_{nl}^{(\chi)}=\frac{k_n^2\sigma_{nl}}{k_l^2-k_n^2}.
\end{equation}
And using Eq. \eqref{eq:apC_relacionbetachichipunto}:
\begin{equation}
    \beta_{nl}^{(\dot{\chi})}=\frac{ik_n\sigma_{nl}}{k_l^2-k_n^2}.
\end{equation}

Summing up, we can write for $n\neq l$:
\begin{equation}
    \beta_{nl}^{(r)}=\begin{cases}
    -\beta_{ln}^{(r)}, &r=L,\Delta L,v\\
    -\beta_{ln}^{(r)}\frac{k_n^2}{k_l^2}, &r=\chi\\
    -\beta_{ln}^{(r)}\frac{k_n}{k_l}, &r=\dot{\chi}
    \end{cases}
    \label{eq:apC_betagenerico}
\end{equation}

\vspace{5 cm}

\bibliographystyle{unsrt}
\bibliography{biblio}

\end{document}